\documentclass[11pt,a4paper]{article}
\usepackage{jheppub}
\usepackage{feynmp}
\usepackage{caption}          
\usepackage{subcaption}       
\usepackage{shuffle}          

\def\be{\begin{equation}}
\def\ee{\end{equation}}
\def\bea{\begin{eqnarray}}
\def\eea{\end{eqnarray}}
\def\beal{\begin{equation}\begin{aligned}}
\def\eeal{\end{aligned}\end{equation}}
\def\nn{\nonumber}

\def\scalz{z}
\def\scalzb{{\bar z}}

\def\spa#1.#2{\left\langle#1\,#2\right\rangle}
\def\spb#1.#2{\left[#1\,#2\right]}

\def\Res_#1{\operatorname*{Res}_{#1}}
\def\Tr{\operatorname*{Tr}}

\def\Trep{T_R}

\def\Tr{\text{Tr}}

\def\eqn#1{eq.~\eqref{#1}}

\def\fig#1{figure~{\ref{#1}}}

\def\sec#1{section~{\ref{#1}}}

\title{Conformal Gravity from Gauge Theory}
\author[a,b,c]{Henrik Johansson,}
\author[b]{Josh Nohle}

\affiliation[a]{Department of Physics and Astronomy, Uppsala University, Box 516, SE-75120 Uppsala, Sweden}
\affiliation[b]{Nordita, KTH Royal Institute of Technology and Stockholm University, Roslagstullsbacken 23, SE-10691 Stockholm, Sweden}
\affiliation[c]{Kavli Institute for Theoretical Physics, University of California, Santa Barbara, CA 93106, USA}

\emailAdd{henrik.johansson@physics.uu.se}
\emailAdd{josh.nohle@gmail.com}

\abstract{We use the duality between color and kinematics to obtain scattering amplitudes in non-minimal conformal ${\cal N}=0,1,2,4$ (super)gravity theories. Generic tree amplitudes can be constructed from a double copy between (super-)Yang-Mills theory and a new gauge theory built entirely out of dimension-six operators. The latter theory is marginal in six dimensions and contains modes with a wrong-sign propagator, echoing the behavior of conformal gravity. The dimension-six Lagrangian is uniquely determined by demanding that its scattering amplitudes obey the color-kinematics duality. The conformal gravity amplitudes obtained from the double copy are compared with the Berkovits-Witten twistor string and shown to agree up to at least eight points in the MHV sector. Our construction can be generalized in a number of ways. Adding scalars to the dimension-six theory gives Maxwell-Weyl gravity, and further adding $\phi^3$ self-interactions among these scalars gives Yang-Mills-Weyl gravity. The latter is identified with Witten's twistor string for maximal ${\cal N}=4$ supersymmetry. Deforming the dimension-six theory by adding a Yang-Mills term, $m^2 F^2$, gives a gauge theory that interpolates between marginal $D=6$ and $D=4$ theories. The corresponding double copy gives an interpolation between conformal gravity and Einstein gravity.}

\preprint{UUITP-21/17 \\
\phantom{~} \hfill NORDITA-2017-060 \\
}

\begin{document}
\maketitle

\pagebreak

\section{Introduction}
\label{sec:intro}

Recent decades have taught us that gauge and gravity theories are intimately connected at a detailed theoretical level. A prime example of this is the {\it double copy} relation between gravity and gauge-theory scattering amplitudes. The double copy can be understood as originating from a duality between color and kinematics~\cite{Bern:2008qj,Bern:2010ue} present in a variety of different gauge theories~\cite{Bern:2008qj,Bern:2010ue,Johansson:2014zca,Johansson:2015oia,Bargheer:2012gv}. 

In gauge theories, such as (super-)Yang-Mills theory or (super-)QCD, color-kinematics duality imposes constraints the detailed form of the amplitudes. At tree level, color-ordered $n$-point gluon amplitudes satisfy so-called BCJ relations~\cite{Bern:2008qj} that can be used to eliminate all but $(n-3)!$ independent partial amplitudes~\cite{Stieberger:2009hq,BjerrumBohr:2010hn}. Similar identities have been worked out~\cite{Johansson:2015oia} and proven~\cite{delaCruz:2015dpa} for QCD with massive quarks.  At loop level, the duality interlocks the kinematic numerators of the individual diagrams entering the amplitude making it possible to determine a majority of the contributions from a small set of master diagrams~\cite{Bern:2010ue,Carrasco:2011mn,Bern:2012uf,Carrasco:2012ca,Bjerrum-Bohr:2013iza,Bern:2013yya,Nohle:2013bfa,Chiodaroli:2013upa,Johansson:2014zca,Mafra:2015mja,Johansson:2015oia,He:2015wgf,Mogull:2015adi, Johansson:2017bfl}. 

In gravity theories, such as Einstein (super-)gravities~\cite{Bern:2010ue,Bern:2011rj,BoucherVeronneau:2011qv,Bern:2012uf}, with or without additional matter~\cite{Carrasco:2012ca,Chiodaroli:2013upa,Johansson:2014zca,Chiodaroli:2015wal,Johansson:2017bfl}, the duality gives a precise prescription for obtaining tree and loop amplitudes by the double copy: replacing the color structure of one gauge theory with the kinematical structure of a second gauge theory~\cite{Bern:2008qj,Bern:2010ue}. At a technical level, this implies that gravity theories can be identified by pairs of gauge theories. 

{\it Color-kinematics duality} has provided a wealth of insight into the computation of tree-~\cite{Bern:2008qj,BjerrumBohr:2009rd,Stieberger:2009hq,BjerrumBohr:2010hn,Mafra:2011kj,Broedel:2012rc,Bargheer:2012gv,Huang:2012wr,Huang:2013kca, Cachazo:2012uq, Cachazo:2012da, Cachazo:2013hca,Cachazo:2013iea, Johansson:2015oia,delaCruz:2015dpa,Mafra:2015vca,Chiodaroli:2015wal,Bjerrum-Bohr:2016axv,Chiodaroli:2017ngp,Du:2017kpo} and loop-level~\cite{Bern:2010ue,Carrasco:2011mn,Bern:2011rj,BoucherVeronneau:2011qv,Boels:2011tp,Bern:2012uf,Bern:2012uc,Carrasco:2012ca,Boels:2013bi,Bjerrum-Bohr:2013iza,Bern:2013yya,Nohle:2013bfa,Chiodaroli:2013upa,Johansson:2014zca,Chiodaroli:2014xia,Mafra:2015mja,He:2015wgf,Mogull:2015adi, Chiodaroli:2015rdg, Luna:2016idw,He:2016mzd,He:2017spx,Johansson:2017bfl} scattering amplitudes, as well as the ultraviolet behavior~\cite{Bern:2012uf,Bern:2012gh,Bern:2013uka,Bern:2013qca,Bern:2014lha}, of these theories. It is beginning to have a non-trivial impact on the formulation of off-shell structures of the theories, guiding the construction of quantities such as form factors~\cite{Boels:2012ew, Boels:2015yna,Yang:2016ear,Boels:2017skl}, classical solutions~\cite{Saotome:2012vy,Monteiro:2014cda,Luna:2015paa,Ridgway:2015fdl,White:2016jzc,Luna:2016due,Cardoso:2016amd,Luna:2016hge,Goldberger:2016iau,Goldberger:2017frp,Adamo:2017nia}, symmetries~\cite{Borsten:2013bp,Anastasiou:2013cya,Anastasiou:2013hba,Anastasiou:2014qba,Anastasiou:2015vba,Chiodaroli:2016jqw,Chiodaroli:2017ngp,Arkani-Hamed:2016rak} and actions~\cite{Bern:2010yg,Monteiro:2011pc,Cheung:2016prv}. 

Color-kinematics duality also features prominently in string-theory scattering amplitudes~\cite{Stieberger:2009hq,BjerrumBohr:2010hn}, and in amplitudes of effective field theories such as the non-linear-sigma model (NLSM), (Dirac-)Born-Infeld, Volkov-Akulov and special galileon theory~\cite{Chen:2013fya,Cheung:2014dqa,Cachazo:2014xea,Du:2016tbc,Cheung:2016prv,Carrasco:2016ldy}. It is well known that string theory provides a geometric (worldsheet) picture of open--closed string relations~\cite{Kawai:1985xq} and monodromy relations~\cite{Stieberger:2009hq,BjerrumBohr:2010hn,Tourkine:2016bak,Hohenegger:2017kqy}. More recently a plethora of new surprising relations connecting string-, field- and effective-theory amplitudes has been discovered. The new string constructions~\cite{Mafra:2011nw,Mafra:2012kh,Broedel:2013tta,Huang:2016tag, Carrasco:2016ldy,Carrasco:2016ygv} exhibit structures that are more akin to field-theory amplitudes and relations, suggesting a close connection to the Lie-algebra picture~\cite{Monteiro:2011pc, Cheung:2016prv} of the kinematic numerators of gauge theories. Conversely, field-theory amplitudes has been reformulated in geometric terms as objects living on worldsheets through the CHY formalism and the (ambi-)twistor string theory. 

The ${{\cal N}=4}$ twistor string theory goes back to the celebrated 2003 work of Witten~\cite{Witten:2003nn}. The single-trace sector of the twistor string reproduces the four dimensional amplitudes of ${{\cal N}=4}$ super-Yang-Mills theory~\cite{Witten:2003nn,Roiban:2004yf}. However, in the multi-trace sector, or at loop level the gauge theory is well-known to be ``contaminated'' by ${{\cal N}=4}$ conformal supergravity. This particular non-mininal form of conformal supergravity theory is expected to exhibit some interesting features, such as electromagnetic duality~\cite{Berkovits:2004jj}, and potentially it is free of conformal anomalies~\cite{Fradkin:1983tg,Berkovits:2004jj}. In the work of Berkovits and Witten~\cite{Berkovits:2004jj}, the conformal supergravity sector was isolated and tree-level amplitudes were computed to any multiplicity in the MHV sector.  

In the current work we will obtain the amplitudes of the Berkovits-Witten conformal gravity through a double copy formulation, involving a new dimension-six gauge theory (in $D=6$ counting). The gauge theory is uniquely pinned down using the duality between color and kinematics. The double-copy construction will generalize to higher-dimensional theories as well as lower-supersymmetric conformal gravities. Furthermore, the construction suggest a number of deformations that can be done on the dimension-six gauge theory, some that goes beyond the context of conformal supergravities, giving interpolations between Einstein gravity and conformal gravity through the double copy. We will also introduce a Yang-Mills sector in the conformal gravity theory; thus the double copy will be able to reproduce the original twistor-string theory of Witten, where ${{\cal N}=4}$ super-Yang-Mills is coupled to ${{\cal N}=4}$ conformal supergravity. The ease of the constructions suggest that loop-level generalizations should be possible, but this topic is beyond the scope of the current paper. 

This paper is organized as follows: in \sec{sec:CK}, we review color-kinematics duality and the double copy in a generalized setting that accommodates higher-derivative gauge theories and conformal gravity.  In \sec{sec:CG_preliminaries}, we introduce the pure conformal Weyl gravity, and we then discuss generic details of matter-coupled, supersymmetric and higher-dimensional conformal gravities. In \sec{sec:6D}, we find the new dimension-six gauge theory through an ansatz procedure. In \sec{deforms}, we consider deformations of the dimension-six gauge theory, and, similarly, deformations of conformal gravity. Conclusions are presented in \sec{sec:conclusion} and explicit Feynman rules for the dimension-six gauge theory is presented in appendix \ref{FeynRules}. In appendix \ref{LieAlgebraSect}, useful Lie-algebra relations are given pictorially.

\section{Color-Kinematics Duality and Tree Amplitudes}
\label{sec:CK}

In this section we give a brief review of color-kinematics duality and the double copy construction. We focus on tree-level amplitudes for external adjoint particles in generic gauge theories (not necessarily Yang-Mills gauge theories), and tree amplitudes in gravities (not necessarily Einstein gravity), which is relevant to the current work. 

A gauge-theory tree-level $m$-point scattering amplitude can generically be written on the form
\be
{\cal A}_m= g^{m-2}\sum_{i \in {\rm cubic}} \frac{n_i c_i}{D_i}\,,
\label{CubicAmp}
\ee
where the sum runs over all cubic tree graphs, of which there are exactly $(2m-5)!!$. The denominators $D_i=\prod p^2_j$ are products of the squared momenta of each internal line of the graph, $c_i$ are the group-theory color-factor for each graph, and $n_i$ are kinematic numerators that are functions of the kinematic data (momenta, polarizations, etc.). Contact interactions, which are normally thought of as higher-valency graphs, are here absorbed into the numerators of the cubic graphs. 

Generically the color factors depend on the details of the group-theory representations and the corresponding Clebsch-Gordan coefficients; however, for the purpose of this paper we assume that they are expressible in terms of $f^{abc}$ structure constants, whose adjoint indices contracted following the connectivity of the cubic graphs. The $c_i$ then obey identities between triplets of graphs $(i,j,k)$, corresponding to Jacobi identities of the structure constants,
\be
c_i+c_j+c_k=0\,.
\label{colorID}
\ee 

The kinematic numerators $n_i$, which are gauge-dependent objects, are typically complicated functions of the kinematic data of the gauge theory making them difficult to write down explicitly. Even if not necessary, in a Yang-Mills theory it is often convenient assume that they are local (polynomial) functions; however, for the higher-derivative gauge theories that we consider in the following sections it is unavoidable that the $n_i$ will have additional poles beyond those manifestly exhibited in the denominator $D_i$ of $\eqn{CubicAmp}$. Indeed, if the kinetic term in the Lagrangian of the gauge theory has more than two derivatives, then these poles cannot be eliminated by a gauge choice as they correspond to states that are present in the theory.  

Even if non-local, in many gauge theories it is possible to find numerators $n_i$ that obey similar identities as the $c_i$ between triplets of graphs $(i,j,k)$
\be
n_i+n_j+n_k=0~~~~~  \Leftrightarrow  ~~~~~  c_i+c_j+c_k=0\,.
\label{kinID}
\ee 
Once these kinematic identities are in a one-to-one correspondence with the color identities (\ref{colorID}), we say that the amplitude exhibits color-kinematics duality. If all amplitudes in a theory can be brought to this form we say that that the theory obeys color-kinematics duality. In practice, this statement is often difficult to prove, in particular for loop amplitudes.  For pure (super-)Yang-Mills gauge theories color-kinematics duality has been proven to  hold (at least) at tree level. 

Stating color-kinematics duality in terms of the numerator identity (\ref{kinID}) is very general, it can be applied to tree and loop amplitudes alike, or even to off-shell correlation functions and Lagrangian interaction terms; however, is not a gauge invariant statement since the numerators of individual diagrams in a gauge theory are not gauge invariant. Fortunately, at tree level, color-kinematics duality can be phrased in terms of gauge invariant relations between different partial tree amplitudes. 

We use the trace-basis decomposition of a gauge-theory tree amplitude with external adjoint particles (in say $SU(N_c)$)
\be
{\cal A}_m= g^{m-2}\sum_{\sigma \in S_{m-1}} {\rm Tr}(T^{a_{1}}T^{a_{\sigma(2)}} \cdots T^{a_{\sigma(m)}}) \, A(1,\sigma(2), \ldots, \sigma(m))
\label{TrDec}
\ee
where the sum runs over $(m-1)!$ permutations and $A(1,\sigma(2), \ldots, \sigma(m))$ are the gauge-invariant color-ordered partial amplitudes. For many familiar gauge theories it is obvious from the Lagrangian that the tree amplitude can be entirely expressed in terms of contracted strings of $f^{abc}$, in which case, through successive use of Jacobi identities, it can be brought to the Del Duca-Dixon-Maltoni (DDM)~\cite{DelDuca:1999rs} form
\be
{\cal A}_m= g^{m-2}\sum_{\sigma \in S_{m-2}} f^{a_1 a_{\sigma(2)} b_1}f^{b_1 a_{\sigma(3)} b_2} \cdots f^{ b_{m-3} a_{\sigma(m-1)} a_m}  \, A(1,\sigma(2), \ldots,\sigma(m-1), m)\,,
\label{DDMDec}
\ee
which is valid for any gauge group. 

The fact that the partial tree amplitudes in \eqn{DDMDec} are the same as those in \eqn{TrDec} is guaranteed by the Kleiss-Kuijf amplitude relations~\cite{Kleiss:1988ne}. The simplest case of Kleiss-Kuijf relations is given by the photon-decoupling identity,
\be
A(1,2,3,\dots, m)+A(2,1,3,\dots, m)+A(2,3,1,\dots, m)+\dots+A(2,3,\dots, 1, m)=0\,.
\label{PhoDec}
\ee
While the photon-decoupling identity can be derived from the statement that $U(1)$ particles do not couple with gluons (i.e. gluons carry no $U(1)$ charge), the more general Kleiss-Kuijf relations are a consequence of the statement that the trees can be entirely expressed in terms of $f^{abc}$, i.e. no other group-theory tensors (such as symmetrized-trace tensors $d^{abc\ldots}$) show up in the amplitude (\ref{DDMDec}).

The gauge-dependent kinematic Jacobi relations $(\ref{kinID})$ can then be translated into gauge-invariant relations for the partial amplitudes. The simplest such BCJ relation takes the form 
\be
p_{1\mu}\, \big[p_2^\mu A(2,1,3,\dots, m)+p_{23}^\mu A(2,3,1,\dots, m)+\dots+p_{23\dots m-1}^\mu A(2,3,\dots, 1, m)\big]=0\,,
\label{BCJrel}
\ee
where $p_{23\ldots n}=p_2+p_3+\ldots+p_n$ are the momenta of the external states. Other BCJ relations can be derived from this one by considering all permutations of the the arguments and simultaneously solving the obtained system of equations. In the end, for adjoint external particles, and for theories that obeys color-kinematics duality, this reduces the independent partial amplitudes to a $(m-3)!$ basis.  

If the BCJ relations (\eqn{BCJrel} and its permutations) hold for all tree amplitudes of a given theory then the theory obeys color-kinematics duality, at least at tree level. In practice, this gives the simplest know approach for checking the validity of color-kinematics duality in a given theory. 

Given a second set of kinematic numerators $\tilde n_i$ (belonging to the same or a different gauge theory) that obeys color-kinematics duality (\ref{kinID}), we can perform the replacement  $c_i \rightarrow \tilde n_i$ in the amplitude (\ref{CubicAmp}) without spoiling gauge invariance. This gives the following double-copy formula
\be
{\cal M}_m= \Big(\frac{\varkappa}{2}\Big)^{m-2}\sum_{i \in {\rm cubic}} \frac{n_i  \tilde n_i}{D_i}
\label{DCamp}\,,
\ee
were we have also replaced the coupling $g \rightarrow \varkappa/2$. Given that the $n_i$ and $\tilde n_i$ contains interactions of spin-1 particles (in two possibly different gauge theories), the double copy will correspond to scattering amplitudes in a theory with interacting spin-2 particles. Furthermore, the (linearized) gauge invariance of the gauge-theory amplitude (\ref{CubicAmp}) is carried through to the double-copy amplitude and it is enhanced to (linearized) diffeomorphism symmetry. This statement follows in general spacetime dimension without specifying the details of the theories or amplitudes that are under consideration.  Note that for \eqn{DCamp} to be valid it is sufficient that one set of the numerators ($n_i$ or $\tilde n_i$) {\it manifestly} obeys the duality (\ref{kinID}). To guarantee diffeomorphism symmetry of the resulting double copy, the other set must also obey color-kinematics duality, but numerators may be calculated in a gauge where this is not manifest.  

The double copy (\ref{DCamp}), thus describes tree amplitudes in a gravitational theory. In the most well-studied cases, when $n_i$ and $\tilde n_i$ comes from two Yang-Mills theories, the double copy will give amplitudes in some Einstein-type gravity theory (i.e. Hilbert-Einstein action plus matter).  It is useful to denote the double copies that can be obtained from various gauge theories by the notation,
\be
\text{gravity th.} = (\text{gauge th.}) \otimes \widetilde{(\text{gauge th.})}
\ee
E.g. some well-known examples are the maximally supersymmetric Einstein and Yang-Mills theories,  $({\cal N}=8~\text{SG}) = ({\cal N}=4~\text{SYM}) \otimes ({\cal N}=4~\text{SYM})$, the half-maximal case $({\cal N}=4~\text{SG}) = ({\cal N}=4~\text{SYM}) \otimes (\text{pure YM})$, or a non-supersymmetric extension of pure Einstein theory $(\text{GR\,+\,dilaton}+B^{\mu\nu}) = (\text{pure YM}) \otimes (\text{pure YM})$, where $B^{\mu\nu}$ can be dualized to an axion in four dimensions.
In this paper, we will consider gauge theories that are not necessarily of the Yang-Mills type, and obtain double copies for gravity theories which are of higher-derivate type, such as conformal Weyl gravities. Before specifying the theories, we conclude this section by giving a few more useful formulas. 

At tree level, the double-copy (\ref{DCamp}) can be phrased in terms of gauge-invariant partial amplitudes. This is the well-known Kawai-Lewellen-Tye formula~\cite{Kawai:1985xq} (in the field-theory limit of its original string-theory context), which we schematically write as
\be
{\cal M}_m =  \Big(\frac{\varkappa}{2}\Big)^{m-2} \sum_{\rho,\sigma \, \in \, S_{m-3}}A(\rho) S(\rho,\sigma) \widetilde{A}(\sigma)\,,
  \label{KLT}
\ee
where $A(\rho)$ and $\widetilde{A}(\sigma)$ are the color-ordered gauge-theory amplitudes, whose arguments are permutations of a $(m-3)!$ BCJ basis, and $S(\sigma,\rho)$ is the KLT kernel which is a polynomial function (or rational function depending on the BCJ basis) of momentum invariants (see e.g. ref.~\cite{BjerrumBohr:2010hn} for explicit formulas for $S(\rho,\sigma)$). The KLT formula gives valid gravity tree amplitudes given that the input are gauge-theory amplitudes for external adjoint particles, and that the latter obeys the BCJ amplitude relations. 

For example, the three-, four- and five-particle KLT relations are
\begin{eqnarray}
M(1,2,3) &\! =\!& i  A(1,2,3) \, \widetilde{A}(1,2,3)\,, \nn\\
   M(1,2,3,4) &\! =\!& - i s_{12} A(1,2,3,4) \, \widetilde{A}(1,2,4,3)\,,  \\
M(1,2,3,4,5) 
&\! = \!& i s_{12} s_{34}  A(1,2,3,4,5) \, \widetilde{A}(2,1,4,3,5) 
+ i s_{13}s_{24} A(1,3,2,4,5) \, \widetilde{A}(3,1,4,2,5) \nn \,,
           \label{KLT45}
\end{eqnarray}
where the gravitational coupling is dropped by defining ${\cal M}_m=(\varkappa/2)^{m-2} M(1,2,\ldots,m)$.

A very useful formula, which can be though of as being half-way between \eqn{DCamp} and \eqn{KLT}, is the double copy written on a DDM form,
\be
{\cal M}_m =  \Big(\frac{\varkappa}{2}\Big)^{m-2} \sum_{\sigma \in S_{m-2}} n_{1\sigma(2)\ldots\sigma(m-1) m}  \, \widetilde{A}(1,\sigma(2), \ldots,\sigma(m-1), m)\,,
\label{DDM_DC}
\ee
which is simply obtained by taking \eqn{DDMDec} and replacing the DDM color factor with a corresponding numerator factor $n_i$, which we label as $n_{1\sigma(2)\ldots\sigma(m-1) m}$. This is a valid operation since the DDM decomposition of the gauge-theory amplitude follow from the Jacobi identity, which the numerators are assumed to satisfy. 

Combining the information from \eqn{KLT} and \eqn{DDM_DC}, and reading off the coefficient of  $\widetilde{A}(\sigma)$, we see that one possible choice for kinematical numerators is~\cite{Kiermaier,BjerrumBohr:2010hn}
\be
n_{1\sigma(2)\ldots\sigma(m-1) m} = \sum_{\rho \, \in \, S_{m-3}}A(\rho) S(\rho,\sigma)\,,
\ee
and the remaining $(2m-5)!!-(m-2)!$ numerators are obtained by applying kinematic Jacobi relations to the above ones. 
This shows that numerators that satisfy color-kinematics duality can always be found at tree level whenever the BCJ amplitude relations are valid. Although this provides an existence proof, in general, the above construction gives non-local numerators (even in Yang-Mills theories) because of the poles in the amplitudes $A(\rho)$; and, furthermore, crossing symmetry of the amplitude is not manifest with these numerators.

\section{Conformal Gravity}
\label{sec:CG_preliminaries}

In this section we discuss some salient features of the conformal supergravities that will later show up in the double copy. However, we start discussing the pure Weyl theory, which is a clean example even if its tree-level S-matrix is trivial in four-dimensional flat space~\cite{Maldacena:2011mk,Adamo:2013tja,Adamo:2016ple}. 
 
\subsection{Pure (Weyl)$^2$ Conformal Gravity}
Pure conformal Weyl gravity in $D=4$ dimensions is a four-derivative theory that consists of a physical spin-2 graviton and associated spin-2 and spin-1 massless ghosts, in total six degrees of freedom packaged into the metric $g_{\mu\nu}$. The action is written in terms of the square of the Weyl tensor,
\be
{\cal S} =\frac{1}{\varkappa^2} \int d^4 x \, \sqrt{-g} \, (W_{\mu\nu\rho\sigma})^2\,,
\label{Waction}
\ee
where $\varkappa$ is a dimensionless coupling constant.
In $D=4$ the relations to standard curvature tensors and invariants are
\bea
W_{\mu\nu\rho\sigma}&=& R_{\mu\nu\rho\sigma}+g_{\nu [\rho} R_{\sigma]\mu}-g_{\mu [\rho} R_{\sigma]\nu}+ \frac{1}{3} g_{\mu [\rho} g_{\sigma]\nu} R \,, \nn \\
(W_{\mu\nu\rho\sigma})^2&=&(R_{\mu\nu\rho\sigma})^2 -2 (R_{\mu\nu})^2+\frac{1}{3} R^2\,,
\eea
where the antisymmetrization $[\cdot \, \cdot]$ includes a standard factor of $1/2$. In $D=4$ dimensions the Gauss-Bonnet term is topological and hence can be added to the action without changing the (classical) theory in a spacetime that is asymptotically Minkowski,
\be
{\rm GB} =(R_{\mu\nu\rho\sigma})^2 -4 (R_{\mu\nu})^2+ R^2 =  (W_{\mu\nu\rho\sigma})^2 -2 (R_{\mu\nu})^2+ \frac{2}{3}R^2 \,.
\ee

The Weyl tensor with one raised index, $W^{\mu}_{\ \ \nu\rho\sigma}$, is invariant under local rescalings of the metric,
\be
g_{\mu\nu}(x) \rightarrow  g_{\mu\nu}(x) \, e^{\phi (x)}\,.
\ee
Under such conformal transformations the Christoffel symbols vary as
\be
\delta \Gamma_{\ \mu \nu}^{\rho}=
\frac{1}{2}( \delta^{\rho}_{\mu}  \, \partial_\nu \phi+ \delta^{\rho}_{\nu} \, \partial_\mu \phi-g_{\mu \nu} \partial^\rho \phi)\,.
\label{varG}
\ee
And the Riemann and Ricci tensors have the variations
\bea
\delta R^\mu_{\ \nu \rho \sigma} &=&  \delta^{\mu}_{[\sigma} \, \nabla_{\rho]} \partial_{\nu} \phi
- g_{\nu [\sigma} \nabla_{\rho]} \partial^\mu \phi + 
\frac{1}{2}  \Big( \delta^{\mu}_{[\rho}   \partial_{\sigma]} \phi \, \partial_\nu \phi 
-g_{\nu [\rho}  \partial_{\sigma]} \phi \, \partial^\mu \phi 
 -\delta^{\mu}_{[\rho}  g_{\sigma] \nu} \partial_\lambda \phi \, \partial^\lambda \phi \Big)\,,  \nn \\  
 \delta R_{\mu \nu} &=&
 -\nabla_{\mu} \partial_{\nu} \phi - \frac{1}{2} g_{\mu \nu}\, \Box \phi 
 +\frac{1}{2} \partial_{\mu} \phi \, \partial_\nu \phi - \frac{1}{2}  g_{ \mu \nu}  \partial_{\lambda} \phi \, \partial^\lambda \phi \,,
\eea
where $\nabla_{\mu}$ is the covariant derivative and $ \Box \phi=\nabla_{\lambda} \partial^\lambda \phi$\,. The Ricci scalar transforms as
\be
R \rightarrow e^{-\phi} \Big(R 
 - 3\, \Box \phi 
 - \frac{3}{2} \partial_{\lambda} \phi \, \partial^\lambda \phi \Big)\,.
\ee
Indeed, plugging in the above transformations into the Weyl tensor, it is straightforward to show that it is invariant $\delta W^{\mu}_{\ \ \nu\rho\sigma}=0$. For the square of the Weyl tensor, the inverse metric factors induce the transformation $(W_{\mu\nu\rho\sigma})^2 \rightarrow e^{-2\phi} (W_{\mu\nu\rho\sigma})^2$, and together with the transformations $\sqrt{-g} \rightarrow e^{2\phi} \sqrt{-g} $ and, the action (\ref{Waction}) is invariant under conformal transformations.

While conformal gravity has aesthetically pleasing features, it is believed to not be a consistent theory because of the ghost states that come with the opposite-sign propagators compared to the physical graviton, and thus indicates a problem with unitarity. To better observe these unwanted signs and states one often deforms the theory by adding an Einstein-Hilbert term, $m^2 R$, to the action whereby the ghost states becomes massive and the $1/p^4$ propagators can be resolved. Ignoring the tensor structure and the overall normalization, the relevant part of the propagator is
\be
\frac{1}{p^4} \rightarrow -\frac{1}{m^2}\Big( \frac{1}{p^2} - \frac{1}{p^2-m^2} \Big)\,,
\label{mGhost}
\ee
which shows the problematic relative sign between the massless and massive state. The problem goes away in the $m\rightarrow \infty$ limit where the massive ghost decouples (giving Einstein gravity), but not obviously so in the conformal gravity limit, $m \rightarrow 0$.
The appearance of the overall $1/m^2$ pole may suggest that the $m \rightarrow 0$ limit is not smooth; however, this is a spurious pole since the difference between the massless and massive propagator vanish as ${\cal O} (m^2)$. Nevertheless, the limit is not quite smooth since the deformed theory has $2+5=7$ states (massless + massive graviton), which becomes six in the $m \rightarrow 0$ limit. The reason for the removal of one state is the Weyl symmetry, or local conformal symmetry, that emerges in this limit. 

When considering supersymmetric or other extensions of conformal gravity the above pole structure will generalize to other particles. States associated with $1/p^4$ poles always come in pairs, or doublets, containing a physical state and a corresponding ghost state. Except of the sign in front of the pole these states have identical quantum numbers (for $m=0$). Further ghost states corresponding to $1/p^2$ poles also appear (such as the spin-1 ghost in the pure theory), and these will not appear in doublets. 

Beyond the details of the states and propagators, the pure conformal gravity S-matrix turns out to not be very interesting to us. The tree amplitudes for physical gravitons vanish in flat four-dimensional space~\cite{Maldacena:2011mk,Adamo:2013tja,Adamo:2016ple}, and thus to have something nontrivial to study we will consider extensions of the pure theory. These extended theories are also what appears in the double copy construction.

\subsection{Matter-Coupled, Non-Minimal Conformal Gravities and $D>4$ Dimensions \label{nonminimal}}

The pure Weyl theory can be generalized in a variety of ways, by adding matter or supersymmetry, by adding non-minimal couplings, by considering higher-dimensional versions, or by introducing a (conformal) Yang-Mills sector. Higher-dimensional extensions of curvature-squared actions will necessarily involve dimensionfull coupling constants, implying that they are not conformal in a strict sense, nevertheless we will refer to them as conformal gravities (as often done in the literature) since they retain some properties of their four-dimensional siblings. 

Before discussing supersymmetric generalizations, we should consider a bosonic extension to the theory that is relevant to the double copy that we later prescribe. We will not give the complete action, but only sketch out some of needed modifications, sufficient to compute three-point amplitudes. Firstly, it is natural to consider the field $\phi(x)$, the {\it dilaton}, to be part of the theory, and hence it should be promoted to an interacting and propagating field by modifying the Weyl action. While this seems to spoil the invariance under scale transformations, one can require that the action remains invariant under simultaneous transformations of the metric and the dilaton. 

Secondly, in addition to the $(W_{\mu\nu\rho\sigma})^2$ curvature invariant, we may use the Levi-Civita tensor to construct the conformal pseudo-invariant
\be
\frac{i}{2} \epsilon_{\rho\sigma \kappa \lambda} \, W^{\mu\nu\rho\sigma} \,W_{\mu\nu}^{\ \ \ \kappa \lambda}\,.
\ee
However, unless we (non-minimally) couple it to some additional field it is a topological term. Therefore let us introduce the pseudo-scalar {\it axion} field, $a(x)$, which we take to behave as an angle. In (extended) Maxwell theory the axion rotates the electromagnetic field strength into the dualized field strength (i.e. electomagnetic duality). The same rotation can be applied to the Weyl tensor,\footnote{One can equivalently rotate the Riemann tensor; however, in Einstein gravity this has no physical effect since the dualized Riemann tensor is traceless and hence it does not contribute in a two-derivative action.}  
\be
W_{\mu\nu\rho\sigma} \rightarrow {\rm cos}(a) W_{\mu\nu\rho\sigma}- i\,{\rm sin}(a)  \widetilde{W}_{\mu\nu\rho\sigma} \,.
\label{WWdual}
\ee
where we defined the dualized Weyl tensor $ \widetilde{W}_{\mu\nu\rho\sigma}=\frac{i}{2} \sqrt{-g} \, \epsilon_{\rho\sigma \kappa \lambda} W_{\mu\nu}^{\ \ \ \kappa \lambda}$, such that the square is $(\widetilde{W}_{\mu\nu\rho\sigma})^2=(W_{\mu\nu\rho\sigma})^2$.

Including a non-minimal coupling to the dilaton generalizes \eqn{WWdual} to a rotation by a complex angle, parametrized by the holomorphic scalar $z= \phi + i a$ and its conjugate, 
\bea
W_{\mu\nu\rho\sigma} \rightarrow W_{\mu\nu\rho\sigma}(\phi, a) &\equiv& e^\phi {\rm cos}(a) W_{\mu\nu\rho\sigma}- i e^\phi {\rm sin}(a)  \widetilde{W}_{\mu\nu\rho\sigma} \, \nn \\
&=& e^{\bar z } \, W_{\mu\nu\rho\sigma}^{+} + e^{ z} \, W _{\mu\nu\rho\sigma}^{-}
\label{WWdual2}
\eea
where $W_{\mu\nu\rho\sigma}^{\pm} = (W_{\mu\nu\rho\sigma} \pm   \widetilde{W}_{\mu\nu\rho\sigma})/2 $ are the selfdual and anti-selfdual Weyl tensors.
The square of the rotated Weyl tensor,
\bea
\Big(W_{\mu\nu\rho\sigma}(\phi, a)\Big)^2&=& e^{2\phi} {\rm cos}(2a)  (W_{\mu\nu\rho\sigma})^2+ e^{2\phi}  {\rm sin}(2a)  \widetilde{W}_{\mu\nu\rho\sigma} W^{\mu\nu\rho\sigma} \nn \\
&=& e^{ 2\bar z } (W_{\mu\nu\rho\sigma}^{+})^2 + e^{ 2z}  (W _{\mu\nu\rho\sigma}^{-})^2 \,,
\label{squaredWWdual}
\eea
is an invariant under the simultaneous transformations,
\be
g_{\mu\nu} \rightarrow  g_{\mu\nu}\, e^{\phi_c (x)} ~~~~  \, \phi \rightarrow \phi+  \phi_c (x) \,, ~~~~ a \rightarrow a +\pi n~~(n\in  \mathbb{Z} )\,,
\ee
where we now switched notation and use $\phi_c$ to denote the classical dilaton field, and $\phi$ for the quantum one.  

An alternative non-minimal coupling that can be introduced, using the same fields, is the following one:
\be
e^{2\phi} {\rm cos}(2a) \, {\rm GB}+  e^{2\phi}  {\rm sin}(2a)  \, \widetilde{W}_{\mu\nu\rho\sigma} W^{\mu\nu\rho\sigma}\,,
\label{NMGB}
\ee
where now the Gauss-Bonnet term is no longer a total derivative because of the fields that multiply it. The interaction terms in \eqn{NMGB} gives exactly the same three-point tree amplitudes as the ones in~\eqn{squaredWWdual}. These non-vanishing amplitudes are in four dimensions given by 
\be
M(1^{--},2^{--},3^{-+}_\scalz)=  -i \spa{1}.{2}^4\,, ~~~~~ M(1^{++},2^{++},3^{+-}_\scalzb)=  -i \spb{1}.{2}^4\,,
\label{3ptAmpW2}
\ee
where the first two particles are gravitons and the third is a holomorphic $z$ or antiholomorphic $\bar z$ scalar, respectively. The somewhat redundant notation using helicity labels for the scalars will become useful when considering the double copies needed to reproduce these amplitudes. 

We will not write down the kinetic terms of the scalars, but we note that they are expected to contain terms of the form $e^{2\bar z}\Box^2 z + e^{2z}\Box^2 \bar z$ giving rise to $1/p^4$ propagators. The (anti-)holomorphic exponential factors are expected to appear by comparing to the Berkovits-Witten twistor string~\cite{Berkovits:2004jj}, which computes tree amplitudes in a non-minimal  ${\cal N}=4$ conformal supergravity theory. The bosonic theory sketched here is better defined as a subsector of the Berkovits-Witten theory. Namely, restrict to those tree amplitudes that are obtained from scattering the top and bottom components (physical scalars and gravitons) of the chiral and anti-chiral ${\cal N}=4$ Weyl multiplets. Further details of the action can be found in ref.~\cite{Berkovits:2004jj}.

${\cal N}=4$ is believed to be the maximal amount of supersymmetry that can coexist with local conformal symmetry in four dimensions~\cite{deWit:1978pd}. The ${\cal N}=4$ conformal supergravity is historically classified into minimal and non-minimal versions, where the terminology comes from Fradkin and Tseytlin~\cite{Fradkin:1983tg,Fradkin:1985am}, who studied conformal anomalies in these theories. The non-minimal case refers to the appearance of scalar functions in front of the Weyl tensor, similar to the terms considered above. In the minimal case these functions are set to unity. It is known that the entire freedom in defining the ${\cal N}=4$ theory can be captured by a single homogeneous function depending on the holomorphic scalars that parametrize the $SU(1, 1)/U(1)$ coset space of the theory. A complete action for ${\cal N}=4$ (non-minimal) conformal supergravity theory has recently been suggested in ref.~\cite{Butter:2016mtk}.

The ${\cal N}=4$ conformal supergravity theory has higher-dimensional uplifts, or extensions, with maximal supersymmetry.
In six dimensions, there seems to exists two distinct ${\cal N}=2$ conformal supergravity actions, one corresponding to the supersymmetrization of the (Riemann)$^2$ tensor and one corresponding to the Gauss-Bonnet term~\cite{Bergshoeff:1986wc,deRoo:1991at}. However, the ten-dimensional ${\cal N}=1$ conformal supergravity theory has a unique action first written down by de Roo~\cite{deRoo:1991at} (see also ref.~\cite{Bergshoeff:1982az} for the linearized theory). The uniqueness property of this theory will agree with the double copy that we obtain; the double copy will use tree amplitudes from ${\cal N}=1$ SYM in ten dimensions and a higher-derivative gauge theory that appears to be unique in higher dimensions. This will give non-trivial evidence in favor of identifying the Berkovits-Witten conformal gravity~\cite{Berkovits:2004jj} with the dimensional reduction of the ten-dimensional theory~\cite{deRoo:1991at}.

Known theories with reduced supersymmetry in four dimensions are the ${\cal N}=1$ \cite{Townsend:1979ki,Kaku:1978nz} and ${\cal N}=2$~\cite{Bergshoeff:1980is} conformal supergravity theories (see also refs.~\cite{Kaku:1977pa,Ferrara:1977mv}). The tree amplitudes for physical gravitons will again vanish in flat four-dimensional space for the minimal theories~\cite{Maldacena:2011mk,Adamo:2013tja,Adamo:2016ple}. Thus we will mainly be interested in versions of these theories that are directly inherited from the Berkovits-Witten twistor string by supersymmetry truncation (again keeping the tree amplitudes given by the top and bottom parts of the ${\cal N}=4$ multiplets). These are non-pure theories as extra dilaton-axion matter multiplets (chiral or hyper) are present in both cases, and they are non-minimal extensions since the dilaton and axion prefactors of the (Weyl)$^2$ term will remain.

Finally, conformal gravity can be coupled to Maxwell or Yang-Mills theory~\cite{Ferrara:1977ij,Kaku:1977rk,Das:1978nr}, which is a particular case of gauged conformal gravity. For simplicity we will refer to these as Maxwell-Weyl and Yang-Mills-Weyl theories. The combined theories enjoys local conformal symmetry making them potentially interesting unified models as they have soft ultraviolet behavior~\cite{Stelle:1976gc} while incorporating both gauge theory (Standard Model) and gravity sectors (possibly spontaneously broken to Einstein gravity~\cite{Kaku:1978ea}). However, the theoretical consistency appears to be spoiled by the conformal gravity ghosts. 

In the case of maximal supersymmetry, the gauge-theory sector ${\cal N}=4$ SYM is well known to be conformal even beyond the classical level. Nevertheless, the minimal ${\cal N}=4$ Yang-Mills-Weyl theory do have a conformal anomaly unless the number of vector multiplets is precisely four~\cite{Fradkin:1983tg}. The non-minimal theories, such as Witten's twistor string~\cite{Witten:2003nn} (${\cal N}=4$ SYM coupled to conformal gravity), may be free of a conformal anomaly~\cite{Fradkin:1985am,Romer:1985yg}, but instead they have a U(1) R-symmetry anomaly (the amplitudes in \eqn{3ptAmpW2} are non-vanishing due to this anomaly).

\subsection{Dimensional Analysis of Double Copy}

We will here perform a straightforward dimensional analysis that will point us towards the gauge theory needed in the double-copy construction of (non-minimal) conformal gravity. 

Although conformal gravity has poles of type $1/p^4$, and the double copy makes manifest only $1/p^2$ poles, we will assume that the double-copy framework is not modified compared to how it works for Einstein gravity. There are several reasons for this ``default'' assumption. On physical grounds we are ultimately interested in finding ways to couple conformal supergravity with lower-derivative theories, such as Einstein gravity, or (conformal) Yang-Mills theory. If this is to work smoothly the double-copy prescription needs to be similar in all cases. On technical grounds, one can easily check that naive generalizations of color-kinematics duality with manifest $1/p^4$ poles will not result in corresponding BCJ relations (suggesting that numerators sitting on top of these propagators are gauge invariant, contradicting the well-known redundancy of gauge and gravity numerators).

The double copy amplitude we are looking for must then be of the form (\ref{DCamp}); e.g. at four points
\be
M^{\rm CG}(1,2,3,4)=\frac{n_s \tilde{n}_s}{s}+\frac{n_t \tilde{n}_t}{t}+\frac{n_u \tilde{n}_u}{u}
\label{nntilde}
\ee 
where we have two types of numerators, $n_i$ and $\tilde n_i$ $(i=s,t,u)$, belonging to two different gauge theories. The Mandelstam invariants are $s=(k_1+k_2)^2$, $t=(k_2+k_3)^2$ and $u=(k_1+k_3)^2$. Coupling constants have been set to unity to not interfere with the dimensional analysis, as we are only interested in counting the number of derivatives. It is simple to realize that in field theories where every operator has exactly $d$ derivatives, $(\partial_\mu)^d$, the resulting tree amplitudes will also be of dimension $d$, independent of the multiplicity and spacetime dimension. For example, in two-derivative Einstein gravity every Feynman diagram will have two more derivatives in the numerator compared to the derivatives (momenta) in the denominator.

Since conformal gravity is a four-derivative theory the tree diagrams should then be of dimension four, implying that the product of four-point numerators $n_i  \tilde{n}_i$ in \eqn{nntilde} must be of dimension six. For general multiplicity $m$, the dimension is obtained by adding the dimension of the $(m-3)$ propagators to the overall dimension,
\be
n_i \tilde{n}_i ~ \sim ~ (\partial_\mu)^{4+2(m-1)}   ~ \sim ~ \frac{(\partial_\mu)^{4(m-2)}}{(\partial_\mu)^{2(m-3)}}\,,
\ee
 In the last step, we rewrote the seemingly local expression into a non-local one by identifying the dependence on the number of cubic vertices $(m-2)$ and the number of propagators $(m-3)$. The derivatives in a field theory either comes from the vertices or from the propagators, and thus on general grounds we expect these combinations to appear. Indeed, we now get a hint that the missing propagator factors of conformal gravity will appear as non-localities in the objects we usually call ``numerators''.\footnote{Color-kinematics duality and the double copy need not numerators to be local; non-local numerators frequently occur in the literature.}

The simplest way to split the derivatives between the numerators is to assume that one copy, say $\tilde{n}_i$, corresponds to a known gauge theory of Yang-Mills type with one-derivative cubic interactions, and the other one, $n_i$, comes from some unknown gauge theory with three-derivative vertices: 
\be
\tilde{n}_i ~ \sim ~  (\partial_\mu)^{m-2}\,,  ~~~~~~~~~ n_i ~ \sim ~ \frac{(\partial_\mu)^{3(m-2)}}{(\partial_\mu)^{2(m-3)}}\,.
\ee
To better understand the properties of the second theory, it is best to do a ``gedanken calculation'' of an $L$-loop $m$-point cubic diagram using the three-derivative vertices, and reinstating the denominator factors of \eqn{CubicAmp}, giving
\be
\frac{n_i}{D_i} ~ \sim ~ \frac{(\partial_\mu)^{3(m-2+2L)}}{(\partial_\mu)^{4(m-3+3L)}} ~ \sim ~  (\partial_\mu)^{6-m-6L}
\ee
The numerical factors in the exponent of the last expression have physical meaning. Let us gedanken integrate this diagram in $D$ spacetime dimensions by hitting it with $\int d^{DL}\ell$ in momentum space. The engineering dimension of the measure can be made to exactly cancel the $L$-dependent term, given that we are in $D=6$ spacetime dimensions. Thus the theory we are looking for is a marginal (or classically conformal) gauge theory in six dimensions. Similarly, the appearance of the  term $(6-m)$ in the exponent implies that the six-dimensional theory is not renormalized beyond six points, since counterterms have to be local $(6-m)\ge0$. This means that at most six-point interactions will appear in the Lagrangian.

\section{The Dimension-Six Gauge Theory}
\label{sec:6D}

 In this section, we work out the Lagrangian of the unknown gauge theory that enters into the conformal gravity double copy. From dimensional analysis of the doubly copy formula, we can deduce that this gauge theory needs to be marginal in $D=6$. 
 
Thus the task is two write down all dimension-six operators and  constrain the dimensionless parameters by imposing color-kinematics duality on the resulting amplitudes.  Starting with the definitions
\bea
F^3 &=& f^{abc} F^{a\, \mu}_{\nu}  F^{b\, \nu}_{\rho} F^{b\, \rho}_{\mu}  \nn \\
D_{\rho} F_{\mu \nu}^a &=& \partial_{\rho} F_{\mu \nu}^a +  g  f^{abc} A_{\rho}^b F_{\mu \nu}^c  \nn \\
F_{\mu \nu}^a &=& \partial_{\mu} A_{\nu}^a-\partial_{\nu} A_{\mu}^a + g  f^{abc} A_{\mu}^b A_{\nu}^c
\label{fieldDef}
\eea
we can write up four different dimension-six parity-invariant\footnote{We are searching for a theory that can be lifted to any dimension, thus we choose to not consider operators involving contractions with a Levi-Civita tensor. Including such terms in, say, $D=4$ or $D=6$ would, however, likely generalize the current construction.} operators that only involve gauge fields, however, due to Bianchi identities only two of them are independent,
\be
\frac{1}{2}(D^{\rho} F^{\mu \nu a})^2 =D^{\rho} F^{\mu \nu a} D_{\mu} F_{\rho \nu}^a  = (D_{\mu} F^{\mu \nu a})^2 - g\, F^3 \,. 
\ee
Choosing to work with the two operators $(D_{\mu} F^{\mu \nu a})^2$ and $F^3$, we can try to write down a linear combination that gives amplitudes that satisfy color-kinematics duality. 

Before looking at explicit amplitudes it is useful to observe that although the operator $(D_{\mu} F^{\mu \nu a})^2$ contains non-linear interactions up to sixth order, the S-matrix corresponding to this operator is trivial. Indeed, the $F^3$ operator is needed for a nontrivial S-matrix, and conversly the $(D_{\mu} F^{\mu \nu a})^2$ operator is needed for there to be a propagator in this theory. The reader might worry that the kinetic term has four derivatives, corresponding to a $1/p^4$ propagator that give rise to ghosts states. But this is not surprising given that we are indirectly trying to construct conformal gravity, which has such propagators and ghost states.

We can now look at the three-point amplitudes which comes entirely from the $F^3$ term. Color-kinematics duality imposes no constraint on three-point amplitudes, but at least we can check that the double copy of this amplitude with a corresponding Yang Mills amplitude looks reasonable.  

It is well known that in four dimensions the identical-helicity amplitudes are the only non-vanishing three-point matrix elements of the $F^3$ operator. For example, up to overall normalization the all-minus amplitude is given by
\be
A^{F^3}(1^-,2^-,3^-) = i \spa{1}.{2}  \spa{2}.{3}  \spa{3}.{1}
\ee
The corresponding conformal gravity amplitude should now be a product between this amplitude and a Yang-Mills amplitude,
\be
A^{\rm YM}(1^-,2^-,3^+) = i\frac{\spa{1}.{2}^4}{\spa{1}.{2}  \spa{2}.{3}  \spa{3}.{1}} \,.
\ee
This gives an non-zero amplitude between two gravitons and the dilaton-axion holomorphic scalar $z=\phi+ia$,
\be
M^{\rm CG}(1^{--},2^{--},3^{-+}_\scalz) =i A^{F^3}(1^-,2^-,3^-) A^{\rm YM}(1^-,2^-,3^+)  =-i \spa{1}.{2}^4\,.
\label{3pthhz}
\ee
Indeed, from the dimension and spin of this amplitude, it is clear that it comes from a four-derivative theory of spin-two fields, consistent with the properties of conformal gravity. It also agrees with the amplitudes obtained in \eqn{3ptAmpW2} for non-minimally coupled Weyl gravity. It it easy to check that this is a valid component amplitude in non-minimal ${\cal N}=4$ conformal supergravity. As can be seen, the U(1) R-symmetry of the minimal theory is broken already by this three-point tree-level amplitude. The U(1) charge can be computed for each gravitational state by taking the difference between the helicity of the corresponding left and right gauge theory state~\cite{Carrasco:2013ypa}. (Thus the gravitons are neutral and $z$ has charge $-1$ in this convention). 

Starting with the four-point amplitude, we perform the exercise of writing down an ansatz for the dimension-six gauge theory Lagrangian,
\be
{\cal L}_{\rm ans}=\frac{1}{2}(D_{\mu} F^{\mu \nu a})^2 + \frac{ 1}{3!} \rho \, g\, F^3
\label{ans}
\ee
where the kinetic term has been canonically normalized and $\rho$ is a number to be determined. 

For helicity configurations $(--++)$ and $(-+-+)$ the four-gluon amplitudes coming from the ansatz (\ref{ans}) are
\bea
A^{{\cal L}_{\rm ans}}(1^-,2^-,3^+,4^+)&=& \frac{i}{8} \rho^2 g^2 \frac{\spa{1}.{2}^2}{\spa{3}.{4}^2} (u-t) \nn \\
A^{{\cal L}_{\rm ans}}(1^-,2^+,3^-,4^+)&=& 0
\label{ansAmp}
\eea
where the amplitude has been color decomposed, i.e. we are giving the coefficient of $\Tr(T^{a_1}T^{a_2}T^{a_3}T^{a_4})$. As mentioned before, the Mandelstam variables are defined as $s=(p_1+p_2)^2$, $t=(p_2+p_3)^2$ and $u=(p_1+p_3)^2$. If the amplitudes obey color-kinematics duality, the following two BCJ relations should hold:
\bea
0 &\stackrel{?}{=} & t A^{{\cal L}_{\rm ans}}(1^-,2^-,3^+,4^+)-u A^{{\cal L}_{\rm ans}}(2^-,1^-,3^+,4^+)= \frac{i}{8} \rho^2 g^2 \frac{\spa{1}.{2}^2}{\spa{3}.{4}^2} s (t - u) \nn \\
0 &\stackrel{?}{=} & u A^{{\cal L}_{\rm ans}}(1^-,3^+,2^-,4^+)- s A^{{\cal L}_{\rm ans}}(1^-,2^-,3^+,4^+)= \frac{i}{8} \rho^2 g^2 \frac{\spa{1}.{2}^2}{\spa{3}.{4}^2} s (t - u)
\label{fail}
\eea
However, with only two operators this exercise has the solution $\rho=0$, which gives a trivial S-matrix. Thus, we have too look for other dimension-six operators to include, which may involve other fields. In particular we should be looking for a field or (non-local) operator that can cancel the pole $ \sim s/\spa{3}.{4}^2=\spb{4}.{3}/\spa{3}.{4}$, which is a simple pole as opposed to the double pole from the gluon propagator. 

Inspired by the offending expressions in \eqn{fail}, is not difficult to see that the two BCJ relations would hold if we added a term to the four-point amplitude of the form
\be
\Delta=-\frac{i}{8} \rho^2 g^2 \frac{\spa{1}.{2}^2 \spb{3}.{4}^2}{s} \,, 
\label{term}
\ee
with the following contributions in each color order:
\bea
&&A(1^-,2^-,3^+,4^+)\equiv A^{{\cal L}_{\rm ans}}(1^-,2^-,3^+,4^+)+\Delta \,, \nn \\
&&A(2^-,1^-,3^+,4^+) \equiv A^{{\cal L}_{\rm ans}}(2^-,1^-,3^+,4^+)+\Delta \,, \nn \\
&&A(1^-,3^+,2^-,4^+) \equiv A^{{\cal L}_{\rm ans}}(1^-,3^+,2^-,4^+)-2\Delta \,.
\eea
This modification is consistent with the Kleiss-Kuijf identity, 
\be
A(1^-,2^-,3^+,4^+)+A(2^-,1^-,3^+,4^+)+A(1^-,3^+,2^-,4^+)=0\,,
\ee
and the left-hand-sides of \eqn{fail} are modified by $ \Delta (t-u) = \Delta(-2u-s)$, which exactly cancels the unwanted terms in the BCJ relations.  

Looking at the reside of the pole in \eqn{term} it is clear that the term we added corresponds to a  propagating scalar with a factorization into two three-point amplitudes corresponding to the operator $ \varphi F^2$. Indeed, in $D=6$ this operator has engineering dimension six, given that the scalar has a regular two-derivative kinetic term, as the pole suggest. 

Given than we have a scalar in the spectrum, there are two more operators of dimension six that we can write down, namely $(D\varphi)^2$ and $\varphi^3$. Thus we expect a Lagrangian of the schematic form $(DF)^2 + F^3
+(D\varphi)^2+\varphi F^2+ \varphi^3$. However, the last two terms would vanish if we are dealing with an adjoint scalar that only couples through the $f^{abc}$ structure constants. The presence of these terms may instead suggest a coupling through the tensor $d_{\rm F}^{abc}=\Tr(\{T^{a},T^{b}\}T^{c})$; however, this cannot explain why an $s$-channel diagram shows up in the color order $A(1^-,3^+,2^-,4^+)$. Furthermore, the scalar cannot be a singlet under the gauge group since then the term in \eqn{term} would not contribute to the single-trace $\Tr(T^{a_1}T^{a_2}T^{a_3}T^{a_4})$ at four points. Nor can the scalar carry any charge or flavor indices not belonging to the gauge group, since it is sourced by the field strength.

We are forced to assume that the scalar is in some unknown representation of the gauge group. Let us denote the indices of this representation by $\alpha, \beta, \gamma, \ldots$, so that the scalar is $\varphi^\alpha$. The new ansatz for the dimension-six gauge theory is now
\be
{\cal L}=   \frac{1}{2}(D_{\mu} F^{a\, \mu \nu})^2+ \frac{1}{3!} \rho g  F^3  +\frac{1}{2}(D_{\mu} \varphi^{\alpha})^2 + \frac{1}{4} \sigma g \,  C^{\alpha ab}  \varphi^{ \alpha}   F_{\mu \nu}^a F^{b\, \mu \nu }  + \frac{1}{3!}  \tau g \, d^{\alpha \beta \gamma}   \varphi^{ \alpha}  \varphi^{ \beta} \varphi^{ \gamma}
\label{newans}
\ee
where $\rho, \sigma, \tau$ are unknown parameters, and $C^{\alpha ab}$ and $d^{\alpha \beta \gamma}$ are group-theory tensors to be determined. From their appearance in the Lagrangian it is clear that we can choose $C^{\alpha ab}$ to be symmetric in its last two indices and similarly choose $d^{\alpha \beta \gamma}$ to be a totally symmetric tensor. The covariant derivative
\be
D_{\mu} \varphi^\alpha = \partial_{\mu} \varphi^\alpha -  i g  (\Trep^{a})^{\alpha \beta} A_{\mu}^a \varphi^\beta   \\
\label{Dvarphi}
\ee
contains one more unknown tensor: $(\Trep^{a})^{\alpha \beta}$. We can think of $(\Trep^{a})^{\alpha \beta}$ as the generator of some real representation ($\varphi$ is real since $F^2$ is real), thus it is antisymmetric in its last two indices. 

Since $C^{\alpha ab}$, $d^{\alpha \beta \gamma}$, $(\Trep^{a})^{\alpha \beta}$ are covariant tensors they must transform correctly under infinitesimal gauge-group rotations. This implies that they satisfy the relations 
\bea
&&(\Trep^{a})^{\alpha \gamma}(\Trep^{b})^{\gamma \beta}-(\Trep^{b})^{\alpha \gamma}(\Trep^{a})^{\gamma \beta}= i f^{abc} (\Trep^{c})^{\alpha \beta}\,,   \label{1stID} \\
&&f^{bae}C^{\alpha ec}+f^{cae}C^{\alpha be}=i(\Trep^{a})^{\alpha \beta}C^{\beta bc}\,, \label{2ndID}  \\
&&(\Trep^{a})^{\alpha \delta}d^{\delta \beta \gamma}+(\Trep^{a})^{\beta \delta}d^{\alpha \delta \gamma}+(\Trep^{a})^{\gamma \delta}d^{\alpha \beta \delta}=0  \label{3rdID} \,.
\eea
See appendix \ref{LieAlgebraSect} for pictorial representation of these identities. 
 
We can now constrain the unknown parameters and tensors by a careful analysis of the tree amplitudes of the dimension-six Lagrangian. Given that we want to constrain the Lagrangian using the BCJ relations for adjoint fields, we need to look at amplitudes that have only external gluons. From inspecting the Feynman diagrams this implies that the $C^{\alpha ab}$, $(\Trep^{a})^{\alpha \beta}$ and $d^{\alpha \beta \gamma}$ tensors are first available in the four-, five- and six-point amplitudes, respectively.\footnote{Note that the $(DF)^2$ and $F^3$ operators contains quartic, quintic and sextic interactions that are needed for restoring gauge invariance of the linearized operators. The fact that the $C^{\alpha ab}$, $(\Trep^{a})^{\alpha \beta}$ and $d^{\alpha \beta \gamma}$ tensors show up at the same orders is suggestive of that they also cancel something unwanted in the $(DF)^2$ and $F^3$ terms.} Indeed, the rule of thumb is that each free $\alpha,\beta,\gamma$ index needs to be soaked up by a pair of gluons before we can study the tensors. From this point of view we can think of $C^{\alpha ab}$ as a Clebsch-Gordan coefficient for some auxiliary ``bi-adjoint'' representation. In the same sense it might be meaningful to think of $\varphi^\alpha$ as an auxiliary field even though it is propagating.\footnote{When a Yang-Mills term is added to the dimension-six Lagrangian the scalar needs to become massive to maintain color-kinematics duality; in the infinite-mass limit the scalar becomes an auxilliary field in Yang Mills theory.}  

In the four-gluon amplitude we find contractions of the type $C^{\alpha ab}C^{\alpha cd}$ that corresponds to an internal scalar. Since we know that these contributions are needed to cancel some pieces of the pure-gluon diagrams, we can assume that this contraction must equal to a tree-level rank-four adjoint tensor. There are exactly two independent such tensors: $f^{ace}f^{ebd}$ and $(c\leftrightarrow d)$. The relative coefficient is uniquely fixed by the symmetries of the $C^{\alpha cd}$ tensor, and the overall coefficient can be absorbed into the free parameter $\rho$ in the Lagrangian. This fixes the relation
\be
C^{\alpha ab}C^{\alpha cd} = f^{ace}f^{edb}+ f^{ade}f^{ecb}\,.
\label{CCreduction}
\ee
Using this relation all four-gluon amplitudes in the Lagrangian (\ref{newans}) can be computed on a color-ordered form.  In particular, the $(--++)$ amplitude that we looked at before now takes the form
\be
A(1^-,2^-,3^+,4^+)= \frac{i}{8}  g^2 \frac{\spa{1}.{2}^2}{\spa{3}.{4}^2} \left( \rho^2 (u-t) - \sigma^2 s \right)
\ee
The spinor prefactor is symmetric under $1\leftrightarrow2$ interchange so only the expression in the parenthesis need to be plugged into the BCJ relation, giving the constraint
\be
0=t\left( \rho^2 (u-t) - \sigma^2 s \right)- (t\leftrightarrow u)= s (t-u) (\rho^2 - \sigma^2)
\ee
Thus we conclude that $\sigma=\pm \rho$. The ambiguity in the sign reflect the fact that the redefinition $\varphi^\alpha \rightarrow - \varphi^\alpha$ in the Lagrangian does not change the value of any pure-gluon amplitude. Without loss of generality we can choose $\sigma=- \rho$, which is more convenient.   

The $(--++)$ and $(-+-+)$ four-gluon amplitudes are now as follows
\bea
A(1^-,2^-,3^+,4^+)&=& \frac{i}{4}   \rho^2  g^2 u \frac{\spa{1}.{2}^2}{\spa{3}.{4}^2}\,,\nn \\
A(1^-,2^+,3^-,4^+)&=& \frac{i}{4}   \rho^2  g^2  u \frac{\spa{1}.{3}^2}{\spa{2}.{4}^2}\,.
\eea

Because we have the $F^3$ operator in the Lagrangian, we also have nonvanishing amplitudes in other helicity sectors. The all-plus amplitude and one-minus amplitude are as follows:\footnote{Note that the amplitudes have contributions at different orders of $\rho$ since the $(D_{\mu} F^{\mu \nu a})^2$ term does give nonvanishing interactions once the $F^3$ term is included. The absence of ${\cal O}(\rho^0)$ contributions is equivalent to the statement that the $(D_{\mu} F^{\mu \nu a})^2$ operator by itself gives a trivial S-matrix.}
\bea
A(1^+,2^+,3^+,4^+)&=& \frac{i}{4}  \rho (4 + 3 \rho) g^2 u \frac{\spb{1}.{2} \spb{3}.{4}}{\spa{1}.{2}\spa{3}.{4}}\,,\nn \\
A(1^-,2^+,3^+,4^+)&=& -\frac{i}{2}  \rho g^2 \spb{2}.{4}^2  \frac{\spa{1}.{2} \spb{2}.{3}}{\spb{1}.{2}\spa{2}.{3}}\,.
\label{allplus}
\eea
All the four-gluon amplitudes satisfy color-kinematics duality for any value of the parameter $\rho$, and in any spacetime dimension $D$.

Turning to the five-gluon amplitudes we encounter tensor contractions of the form $C^{\alpha ab}(\Trep^{c})^{\alpha \beta}C^{\beta de}$, corresponding to an internal scalar that radiates off a gluon. This tensor contraction is entirely determined by the infinitesimal-transformation identity in \eqn{2ndID}, and the previously determined identity in \eqn{CCreduction},
\be
C^{\alpha ab}(\Trep^{c})^{\alpha \beta}C^{\beta de} = (f^{ahg}f^{gbe}+f^{aeg}f^{gbh}) f^{hcd} +(d \leftrightarrow e) \,.
\label{CTCreduction}
\ee

Similar to the behavior of the four-point all-plus amplitude and one-minus amplitude in \eqn{allplus}, the five-gluon amplitude have separate contributions of ${\cal O}(\rho)$, ${\cal O}(\rho^2)$ and ${\cal O}(\rho^3)$. But unlike before, these contributions do not separately satisfy the five-point BCJ relation
\be
0= 
 s_{25} A(1, 2, 5, 3, 4) + (s_{25} + s_{35}) A(1, 2, 3, 5, 4) +(s_{25}+s_{35}+s_{45})A(1, 2, 3, 4, 5)\,.
\ee 
We find that in order for this relation to hold a unique value for the $\rho$ parameter is selected, ($\sigma$ is also fixed through previous relation):
\be
\rho =-2\,,~~~ \sigma =2.
\ee
With this choice all five-gluon amplitudes (in any dimension) satisfy the BCJ relations, and hence color-kinematics duality is present up to this order. (See \sec{sec:Explicit} for the explicit MHV five-point amplitude.)

Finally, we analyse the six-gluon amplitudes. In these amplitudes we find the tensor contraction between one $d^{\alpha \beta \gamma}$ tensor and three $C^{\alpha a b}$ tensors. We cannot use any of the previously known group-theory relations to reduce this to structure constants $f^{abc}$. Instead we have to rely on the six-point BCJ relations to constrain this tensor structure. By the same logic as before, we can assume that the purpose of this tensor structure is to cancel certain contributions in the pure-gluon Feynman diagrams, and hence the tensor should be a linear combination of tree-level rank-six adjoint-index tensors, which can be constructed out of contracting four $f^{abc}$ tensors. There are $4!=24$ independent such tensors, so in principle we can write down an ansatz and constrain the 24 coefficients. 

However, this would not be a satisfactory solution to the tensor-reduction problem. The reason is that this gives us no information on what we should do when we have several $d^{\alpha \beta \gamma}$ tensors that are contracted with each other, as we will encounter starting with the eight-gluon amplitude. In order for this problem to have a satisfactory solution we must demand that we can reduce a contraction of one $d^{\alpha \beta \gamma}$ tensor and two $C^{\alpha a b}$ tensors, so that we have at least one free $\alpha$-index. This free index can then be used to recursively define all unknown tree-level tensors (in analogy to how a current $J^\mu$ can be used to recursively define all tree amplitudes in Yang-Mills theory). Thus we write down the following general reduction ansatz,
\bea
d^{\alpha \beta \gamma}C^{\beta a b}C^{\gamma cd}&=& x \Big(C^{\alpha a e}(f^{ecg}f^{gdb}+f^{edg}f^{gcb})+ C^{\alpha b e}(f^{ecg}f^{gda}+f^{edg}f^{gca}) \nn \\
&&\null +C^{\alpha c e}(f^{eag}f^{gbd}+f^{ebg}f^{gad})+C^{\alpha d e}(f^{eag}f^{gbc}+f^{ebg}f^{gac})\Big)  \nn \\ && \null + y \, C^{\alpha eg}(f^{eac}f^{gbd}+f^{ead}f^{gbc})
\label{dCCreduction}
\eea
The right-hand-side must involve at least one $C^{\alpha ab}$ tensor, since we have a free $\alpha$ index, but the remaning tensors should be of $f^{abc}$-type given that the reduction is complete. Once the final $\alpha$ index is contracted into another $C^{\alpha ab}$ tensor one can use \eqn{CCreduction} in order to obtain only tree-level tensors for adjoint fields. The ansatz has two parameters $x$ and $y$; however, it is clear that an overall scale can be absorbed into the definition of $d^{\alpha \beta \gamma}$, or equivalently into the unknown $\tau$ that multiplies $d^{\alpha \beta \gamma}$ in the Lagrangian. We can fix this freedom by demanding that $\tau=1$.

Computing the six-point amplitude and imposing the BCJ relation
\bea
0 &=& 
 s_{26} A(1, 2, 6, 3, 4, 5) + (s_{26} + s_{36}) A(1, 2, 3, 6, 4, 5) +(s_{26}+s_{36}+s_{46})A(1, 2, 3, 4, 6,5) \nn \\
 &&\null +(s_{26}+s_{36}+s_{46}+s_{56})A(1, 2, 3, 4, 5,6)\,,
\eea
we find that there is a unique solution for the tensor reduction ansatz,
\be
x=1\,,~~~~y=2\, ~~~~({\rm and}~~ \tau=1) \,.
\ee
And thus all free parameters in the Lagrangian have been fixed, except for the (overall) coupling constant $g$. 

We can rewrite the reduction formula \eqn{dCCreduction} into slightly more compact notation by noting that some of the $f^{abc}$ combinations in brackets are familiar from before in terms of two $C^{\alpha ab}$ tensors, we thus have  
\bea
d^{\alpha \beta \gamma}C^{\beta a b}C^{\gamma cd}&=& C^{\alpha a e} C^{\delta eb}C^{\delta cd} +C^{\alpha b e}C^{\delta ea}C^{\delta cd}+C^{\alpha c e} C^{\delta ed}C^{\delta ab}+C^{\alpha d e} C^{\delta ec}C^{\delta ab}  \nn \\ && \null +2 \,C^{\alpha eg}(f^{eac}f^{gbd}+f^{ead}f^{gbc})\,.
\label{dCCreduction2}
\eea
But more interestingly, using the reduction of a product of three tensors and previously obtained relations we can show that there in fact exists a simpler reduction relation for a product of two tensors only
\be
C^{\alpha ab}d^{\alpha \beta \gamma}=(\Trep^a)^{\beta \alpha} (\Trep^b)^{\alpha \gamma} +(\Trep^b)^{\beta \alpha} (\Trep^a)^{\alpha \gamma} +C^{\beta ac} C^{\gamma cb}+C^{\beta bc} C^{\gamma ca}
\ee
This can be shown be after soaking up the free $\beta$-index with a $C^{\beta cd}$ tensor, and using the previous identities. This new identity is both simpler and fully equivalent to eqs. (\ref{dCCreduction}) and (\ref{dCCreduction2}), so in the following we take this identity as the defining one. 

\subsection{The Dimension-Six Lagrangian}
\label{dim6Lsec}

Let us summarize what we achieved: imposing color-kinematics duality (BCJ relations) up to six points gives a unique\footnote{By uniqueness we mean that all the fudge factors in the Lagrangian ansatz are fixed. Further adding (lower-dimensional) operators or matter will give families of theories satisfying color-kinematics duality.} Lagrangian for the dimension-six theory.  It reads
\be
{\cal L}_{(DF)^2}=   \frac{1}{2}(D_{\mu} F^{a\, \mu \nu})^2- \frac{1}{3} g  F^3  +\frac{1}{2}(D_{\mu} \varphi^{\alpha})^2 + \frac{1}{2}g \,  C^{\alpha ab}  \varphi^{ \alpha}   F_{\mu \nu}^a F^{b\, \mu \nu }  + \frac{1}{3!} g \, d^{\alpha \beta \gamma}   \varphi^{ \alpha}  \varphi^{ \beta} \varphi^{ \gamma}
\label{FullLagrDF}
\ee
where $F^3$ and the covariant derivatives are defined in eqs. (\ref{fieldDef}) and  (\ref{Dvarphi}). The scalar $\varphi^\alpha$ transforms in a real (auxilliary) representation of the gauge group, with generator $(\Trep^a)^{\alpha \beta}$ and symmetric Clebsh-Gordan coefficients $C^{\alpha ab}$ and $d^{\alpha \beta \gamma}$, which are implicitly defined through the two relations
\bea
&&C^{\alpha ab}C^{\alpha cd} = f^{ace}f^{edb}+ f^{ade}f^{ecb}\,, \label{Id1a} \\
&&C^{\alpha ab}d^{\alpha \beta \gamma}= (\Trep^a)^{\beta \alpha} (\Trep^b)^{\alpha \gamma}+ C^{\beta ac} C^{\gamma cb} + (a \leftrightarrow b)\,, \label{Id1b} 
\eea
where the last equation is (non-trivially) equivalent to the tensor reduction in \eqn{dCCreduction}. Furthermore, all the group-theory tensors transforms covariantly under infinitesimal group rotations, implying that they satisfy the three relations in eqs. (\ref{1stID})--(\ref{3rdID}). These five tensor relations are sufficient to reduce any color structure appearing in a pure-gluon tree amplitude to contractions of $f^{abc}$ tensors. See appendix \ref{LieAlgebraSect} for pictorial representation of the identities. 

Using the Feynman rules given in appendix \ref{FeynRules} we have computed the tree-level pure-gluon amplitudes up to multiplicity eight in $D$ dimensions, and found that no additional corrections to the Lagrangian are needed for it to satisfy color-kinematics duality. A curious feature of these amplitudes is that the product between any two polarization vectors, $\varepsilon_i \cdot \varepsilon_j$, will always cancel out through non-trivial identities. This has an interesting consequence when considering dimensional reduction: the scalars that come from the extra-dimensional gluon components will automatically decouple from the theory. Thus, in this theory, dimensional reduction is the same operation as dimensional truncation.

\subsection{Double Copy for Conformal  Weyl  Gravity}
\label{sec:CG}

Starting with two gauge theories that obey color-kinematics duality, the double copy guarantees that we obtain amplitudes containing spin-2 particles and that are invariant under diffeomorphism symmetry~\cite{Chiodaroli:2017ngp}. The resulting amplitude must then come from some gravity theory. A natural double copy to consider is the following construction
\be
{\rm CG}=\big((DF)^2~{\rm theory}\big)\otimes {\rm YM}\,.
\label{CG_DC}
\ee
The gauge theory on the left is the dimension-six theory (\ref{FullLagrDF}); we use the shorthand $(DF)^2$ to denote it. The gauge theory on the right is, in the simplest case, pure Yang-Mills theory. By dimensional analysis, and by construction, the resulting gravity theory is of the four-derivative type. Hence it must be built out of curvature squared invariants: (Weyl)$^2$, Gauss-Bonnet or $R^2$; and possibly linear curvature invariants with two derivatives: $\Box R$, $\nabla^\mu \nabla^\nu R_{\mu \nu}$. We will refer to it as a conformal gravity (CG) theory although the precise action is yet to be determined. Note that the double copy (\ref{CG_DC}) should give valid gravity amplitudes in any spacetime dimension, since both theories are bosonic and obey color kinematics duality in any dimension.

In addition to the regular graviton states, the physical on-shell spectrum of the double copy (\ref{CG_DC}) includes a dilaton and antisymmetric two-form obtained by tensoring the asymptotic vectors of each gauge theory:
\be
A^{\mu} \otimes A^{\nu}\, \sim \, h^{\mu \nu} \oplus B^{\mu \nu} \oplus \phi
\label{hBphi}
\ee
where the Lorentz indices should be projected to little group-indices for a more precise relation. In four dimensions the two-form is equivalent on shell  to the pseudo-scalar axion, introduced earlier.
The states in \eqn{hBphi} correspond to $1/p^4$-type propagator poles and thus they always appear in doublets together with companion ghost states as explained in \sec{sec:CG_preliminaries}. In addition, there are further ghost states that only have $1/p^2$ poles. These are somewhat difficult to isolate in an amplitude since they can mix with contact-like terms coming from the partial cancelations of the double poles $\sim p^2/p^4$.  States originating from the scalar in the $(DF)^2$ theory also come with $1/p^2$ poles, suggesting that these states may combine with the ghosts. The precise details of the states with $1/p^2$ poles, and how to isolate them in the amplitude, we leave for future work.

While we have not proven that the left-hand-side of \eqn{FullLagrDF} is a theory with conformal symmetry, we will see that it gives scattering amplitudes that agree with those of a bosonic sector of the Berkovits-Witten conformal gravity twistor string. 

\subsection{Double Copy for ${\cal N}=1,2,4$ Conformal Supergravities}
\label{sec:CSG}

The double copy in \eqn{CG_DC} can be extended to include supersymmetry. While the dimension-six theory cannot be made supersymmetric due to the appearance of the $F^3$ operator, the left gauge-theory copy can be replaced by supersymmetric Yang-Mills theory with ${\cal N}=1$, ${\cal N}=2$ or ${\cal N}=4$. This gives three different conformal supergravities,
\be
\big({\cal N}=1,2,4~{\rm CSG}\big)=\big((DF)^2~{\rm theory}\big)\otimes \big({\cal N}=1,2,4~{\rm SYM}\big)\,.
\label{CG_DCsusy}
\ee
Similar to the bosonic case, the ${\cal N}=1$ and ${\cal N}=2$ conformal supergravities contain extra states compared to the pure versions of the theories. These are supersymmetric extensions of the dilaton and axion states that appeared in \eqn{hBphi}. Specifically, for ${\cal N}=1$ this gives an extra chiral multiplet, and for ${\cal N}=2$ an extra hypermultiplet, both corresponding to $1/p^4$ poles. Associated ghost states will also appear due to these poles.  

In the case of ${\cal N}=4$ supersymmetry, the double copy (\ref{CG_DCsusy}) gives the correct physical particle content: the dilaton and axion are now part of the ${\cal N}=4$ graviton multiplets. The interaction are of the non-minimal type: the U(1) R-symmetry of the minimal theory is broken already by the three-point tree-level amplitudes. 

In \sec{sec:Explicit}, we compute tree-level amplitudes up to eight external particles and show that it perfectly matches with the Berkovits-Witten MHV formula for non-minimal conformal gravity. As is well known, this particular conformal gravity shows up in the multi-trace sectors of Witten's twistor string theory~\cite{Witten:2003nn}, originally contaminating the pure ${\cal N}=4$ gauge theory, but later on this gravitational sector was considered by itself in the work of Berkovits and Witten~\cite{Berkovits:2004jj}. By supersymmetric truncation of the tree-level amplitudes one can obtain the corresponding tree amplitudes of ${\cal N}=0,1,2$ non-minimal conformal (super)gravities.

By virtue of color-kinematics duality being dimension agnostic, the double copy (\ref{CG_DCsusy}) should also give valid gravity amplitudes in dimensions $D>4$, given that the SYM theory exists. For half-maximal supersymmetry the double copy lifts up to six dimensions giving amplitudes in $D=6$ ${\cal N}=(1,0)$ conformal supergravity~\cite{Bergshoeff:1986wc}. Likewise, for maximal supersymmetry the double copy lifts up to ten dimensions giving amplitudes in $D=10$ ${\cal N}=(1,0)$ conformal supergravity~\cite{deRoo:1991at}.\footnote{Although it is beyond the scope of the current paper, we note that since the mentioned theories are chiral, gravitational anomalies may appear  at loop level.} Since the action of the ten-dimensional theory is given by a unique supersymmetrization of the curvature-squared invariants~\cite{deRoo:1991at}, and the double copy gives supersymmetric curvature-squared amplitudes in ten dimensions, the double copy can be identified with the theory in ref.~\cite{deRoo:1991at}. Lower-dimensional and lower-supersymmetry double copies should then give amplitudes in theories that are related to the ten-dimensional ``parent'' theory by dimensional reduction and supersymmetry truncation.

\subsection{Explicit Amplitudes for $(DF)^2$ and CG}
\label{sec:Explicit}

Using the Feynman rules for the dimension-six gauge theory (see appendix \ref{FeynRules}), we have computed all tree-amplitudes up to eight points, and showed that they obey color-kinematics duality in any spacetime dimension. In this section we will present some of the results at low multiplicity and in four dimensions, and perform the double copy with corresponding Yang-Mills amplitudes.

The color-ordered three and four-gluon amplitudes in the dimension-six theory, denoted by $(DF)^2$, reads
\begin{align}
A^{(DF)^2}(1^-,2^-,3^-)&= i \spa{1}.{2}\spa{2}.{3}\spa{3}.{1}\,,\nn \\
A^{(DF)^2}(1^-,2^-,3^+,4^+)&=- i u \frac{\spa{1}.{2}^2}{\spa{3}.{4}^2}\,,\nn \\
A^{(DF)^2}(1^-,2^+,3^-,4^+)&= -i  u \frac{\spa{1}.{3}^2}{\spa{2}.{4}^2}\,, \nn \\
A^{(DF)^2}(1^+,2^+,3^+,4^+)&=- i u \frac{\spb{1}.{2} \spb{3}.{4}}{\spa{1}.{2}\spa{3}.{4}}\,,\nn \\
A^{(DF)^2}(1^-,2^+,3^+,4^+)&= -i  \spb{2}.{4}^2  \frac{\spa{1}.{2} \spb{2}.{3}}{\spb{1}.{2}\spa{2}.{3}}\,,
\end{align}
where $\pm$ are the helicities of the gluons.  The all-minus and one-plus amplitudes can be obtained by complex conjugation of the above expressions. 
The four-point amplitudes satisfy the Kleiss-Kuijf and BCJ relations, {\it e.g.}
\be
A(1, 2, 4, 3)= \frac{s_{14}}{s_{24}} A(1, 2, 3, 4)\,.
\ee

For later use, we give the MHV gluon amplitudes in Yang-Mills theory,
\be
A^{\rm YM}(1^+,2^+, \ldots, i^-, \ldots, j^-,\ldots, n^+)= i \frac{\spa{i}.{j}^4}{\spa{1}.{2}\spa{2}.{3} \cdots \spa{n}.{1}}\,,
\ee
and in ${\cal N}=4$ super-Yang-Mills theory the MHV superamplitude is
\be
A^{\rm SYM}(1,2, \ldots, n)= i \frac{\delta^{(8)}(Q)}{\spa{1}.{2}\spa{2}.{3} \cdots \spa{n}.{1}}\,,
\ee
where the total supermomentum is $Q^{\alpha A}=\sum_{i=1}^n \lambda_i^\alpha \eta^A_i$.

The conformal gravity amplitudes are given by double copies between the dimension-six theory and Yang-Mills theory. The only non-zero three-point amplitude, in four dimensions, is the graviton-graviton-scalar amplitude that we already gave in eqs. (\ref{3ptAmpW2}) and (\ref{3pthhz}),
\be
M^{\rm CG}(1^{--},2^{--},3^{-+}_\scalz) = i A^{(DF)^2}(1^-,2^-,3^-) A^{\rm YM}(1^-,2^-,3^+) =- i \spa{1}.{2}^4\, .
\ee
The pure-graviton four-point amplitude is
\begin{align}
M^{\rm CG}(1^{--},2^{--},3^{++},4^{++}) &= -i  \frac{st}{u}A^{(DF)^2}(1^-,2^-,3^+,4^+) A^{\rm YM}(1^-,2^-,3^+,4^+) \nn \\
&= i \spa{1}.{2}^4 \frac{ \spb{3}.{4}^2}{\spa{3}.{4}^2}=  i  \frac{\spa{1}.{2}^4 \spb{3}.{4}^4}{s^2}\,,
\end{align}
Note that the gravitational coupling has been suppressed, it can be restored for a $n$-point tree amplitude as follows:
\be
{\cal M}_m=\Big(\frac{\varkappa}{2}\Big)^{m-2} M_m\,,
\ee
where $\varkappa$ is a dimensionless coupling constant in four dimensions.

We can also compute some scalar-graviton amplitudes 
\begin{align}
M^{\rm CG}(1^{+-}_\scalzb,2^{+-}_\scalzb,3^{++},4^{++})& = -i  \frac{st}{u} A^{(DF)^2}(1^+,2^+,3^+,4^+)A^{\rm YM}(1^-,2^-,3^+,4^+) =  i \spb{3}.{4}^4 \,,\nn \\
M^{\rm CG}(1^{--},2^{+-}_\scalzb,3^{++},4^{++})&= -i  \frac{st}{u}  A^{(DF)^2}(1^-,2^+,3^+,4^+)A^{\rm YM}(1^-,2^-,3^+,4^+) \nn \\
&= -i   \spa{1}.{2}^4 \frac{\spb{2}.{3} \spb{3}.{4} \spb{4}.{2} }{\spa{2}.{3}\spa{3}.{4}\spa{4}.{2}}\,, \nn \\
M^{\rm CG}(1^{-+}_\scalz,2^{++},3^{+-}_\scalzb,4^{+-}_\scalzb)&= -i  \frac{st}{u}  A^{(DF)^2}(1^-,2^+,3^+,4^+)A^{\rm YM}(1^+,2^+,3^-,4^-) \nn \\
&= i  s \spa{3}.{4}^2  \frac{ \spb{2}.{3} \spb{2}.{4}}{\spa{2}.{3}\spa{2}.{4}} \,, \nn \\
M^{\rm CG}(1^{-+}_\scalz,2^{-+}_\scalz,3^{+-}_\scalzb,4^{+-}_\scalzb)&= -i  \frac{st}{u}  A^{(DF)^2}(1^-,2^-,3^+,4^+)A^{\rm YM}(1^+,2^+,3^-,4^-) = i s^2  \,, \nn \\
M^{\rm CG}(1^{-+}_\scalz,2^{--},3^{+-}_\scalzb,4^{++})&= -i  \frac{st}{u}  A^{(DF)^2}(1^-,2^-,3^+,4^+)A^{\rm YM}(1^+,2^-,3^-,4^+) = i  \frac{\spa{2}.{3}^4 \spb{3}.{4}^4}{s^2} \,,
\end{align}
Note that these scalars are not part of minimal/pure Weyl gravity; instead they appear in the non-minimal conformal gravity that can be obtained form a supersymmetric truncation of non-minimal ${\cal N}=4$ superconformal gravity, as discussed in \sec{nonminimal}.

We can obtain the ${\cal N}=4$ conformal supergravity amplitudes by double copying the dimension-six theory with ${\cal N}=4$ super-Yang-Mills,
\be
M^{\rm CSG}(1,2,3,4) =  -i \frac{st}{u}A^{(DF)^2}(1,2,3,4) A^{\rm SYM}(1,2,3,4) \,.
\ee
The states assemble into a chiral $(-)$ and antichiral $(+)$ ${\cal N}=4$ graviton multiplet, and in terms of these we have the following superamplitudes in conformal supergravity:
\begin{align}
M^{\rm CSG}(1^-,2^-,3^+,4^+)&= i  \delta^{(8)}(Q) \frac{\spb{3}.{4}^2}{\spa{3}.{4}^2} \,,\nn \\
M^{\rm CSG}(1^+,2^+,3^+,4^+)&= i \delta^{(8)}(Q) \frac{\spb{1}.{2}^2 \spb{3}.{4}^2}{\spa{1}.{2}^2\spa{3}.{4}^2}  \,,\nn \\
M^{\rm CSG}(1^-,2^+,3^+,4^+)&= -i \delta^{(8)}(Q)   \frac{ \spb{2}.{3}\spb{3}.{4}  \spb{4}.{2} }{\spa{2}.{3} \spa{3}.{4} \spa{4}.{2} } \,,
\end{align}
plus amplitudes that are related to these through conjugation.
Note that the chirality $(\pm)$ of the graviton multiplet translate directly to the helicity of the gluons in the dimension-six theory. The nonvanishing of the all-chiral and one-chial amplitudes is thus inherited from the nonvanishing of the all-plus and one-minus amplitudes in the dimension-six theory.

For the five-point amplitude in the dimension-six theory the all-plus and one-minus amplitudes are again non-zero. Unfortunately they are rather lengthy, so we give only the MHV amplitude here. The MHV amplitude, with two neighboring negative helicities, is on the form
\be
A^{(DF)^2}(1^-, 2^-, 3^+, 4^+, 5^+) = -A^{\rm YM}(1^-, 2^-, 3^+, 4^+, 5^+) \big(a + i  \, b \, \epsilon(1,2,3,4)\big)\,,
\ee
where $\epsilon(1,2,3,4)\equiv {\rm Det}(p_1,p_2,p_3,p_4)$. We have used the Yang-Mills amplitude to soak up the helicity weights and thus we can write it in terms of two momentum-dependent functions $a$ and $b$ using only Lorentz dot products. The first function is
\begin{align}
 \! a \! &= \! \frac{(s_{24} + s_{34}) (s_{12} (s_{24} + s_{34}) + (s_{23} - 3 s_{15}) s_{45})}{s_{34} s_{35}} 
+ \frac{(s_{24} + s_{34}) (s_{12} (s_{24} + s_{34}) - (s_{15} - 3 s_{23}) s_{34})}{s_{35} s_{45}} 
\nn \\
& \null
- \frac{s_{34} s_{45} (s_{34} + s_{35}) (s_{15} (s_{34} + s_{35}) + 2 s_{34} s_{45})}{s_{12}^3 s_{23}} 
- \frac{s_{34} s_{45} (s_{35} + s_{45}) (s_{23} (s_{35} + s_{45}) + 2 s_{34} s_{45})}{s_{12}^3 s_{15}} 
\nn \\
& \null
+ \frac{s_{15} (3 s_{15} - 2 s_{23}) s_{45}^2}{s_{12} s_{34} s_{35}} 
+ \frac{s_{23} (3 s_{23} - 2 s_{15}) s_{34}^2}{s_{12} s_{35} s_{45}} 
+ \frac{2 (s_{23} - s_{24}) s_{34} - s_{24}^2}{s_{35}} 
- \frac{s_{23}^2 s_{34}^3}{s_{12}^2 s_{35} s_{45}} 
- \frac{s_{15}^2 s_{45}^3}{s_{12}^2 s_{34} s_{35}} 
 \nn \\
& \null
- \frac{s_{34}^3 s_{45}^3}{s_{12}^3 s_{15} s_{23}}
+ \frac{2 s_{34} (s_{23} (s_{23} + s_{24}) + s_{24} s_{34})}{s_{12} s_{35}} 
+ \frac{s_{34}^2 (s_{34} (2 s_{23} + s_{34}) - 2 s_{15} (s_{23} + s_{34}))}{s_{12}^2 s_{35}} 
\nn \\
& \null
+\frac{s_{13}^2 s_{34}^2 - 2 s_{13} s_{15} s_{34} s_{45} + (s_{15}^2 - 2 s_{34}^2) s_{45}^2}{s_{12}^3} 
- \frac{2 s_{13} s_{15} + s_{34}^2 + (s_{34} - s_{13} - 3 s_{15}) s_{45} + s_{45}^2}{s_{12}} 
\nn \\
& \null
+ \frac{s_{34} (s_{34} (s_{34} - 2 s_{15}) + s_{13} (s_{15} + s_{34})) - s_{15} (s_{13} + s_{34}) s_{45}}{s_{12}^2} + 2 s_{45} + s_{14}\,,
\end{align}
and the second one is
\begin{align}
\! b& \! =\! \frac{(s_{15} + s_{25}) (s_{13} s_{24} - s_{14} s_{23})}{s_{12}^3 s_{34}} 
+ \frac{(s_{14} + s_{24}) (s_{15} s_{23} - s_{13} s_{25})}{s_{12}^3 s_{35}} 
+  \frac{(s_{13} + s_{23}) (s_{14} s_{25} - s_{15} s_{24})}{s_{12}^3 s_{45}}  
\nn \\
& \null
+ \frac{s_{12}^2 - s_{23} (2 s_{15} + s_{25}) - s_{13} (s_{15} - 3 s_{34}) 
+ 3 s_{25} s_{45}}{s_{12}^3} - \frac{s_{34}^2 s_{45}^2}{s_{12}^3 s_{15} s_{23}} 
+ \frac{s_{13} s_{34} s_{45}}{s_{12}^3 s_{23}} + \frac{s_{25} s_{34} s_{45}}{s_{12}^3 s_{15}} \,.
\nn \\
\end{align}
Note that the poles in $a$ and $b$ are in addition to the poles in the $A^{\rm YM}$ factor. Thus we can see that propagators are typically of the type $ \sim 1/s_{ij}^2$ consistent with the four-derivative kinetic term. As is obvious, there are even some poles $1/s_{12}^3$; however, these are spurious poles. They are artefacts of the helicity choice and the fact that we pulled out a $A^{\rm YM}$ factor. The $D$-dimensional amplitude has at most $1/s_{ij}^2$ poles.

Note that the $A^{(DF)^2}(1^-, 3^+, 2^-, 4^+, 5^+)$ amplitude is not related to the above one by a naive permutation of the legs in the $A^{\rm YM}(1^-, 2^-, 3^+, 4^+, 5^+)$ factor. This is because there is no supersymmetry in this theory, and thus the MHV amplitudes with different color orders are not related by supersymmetry Ward identities.  Nonetheless, the different MHV amplitudes are still related though the BCJ relations, implying that we can straightforwardly calculate the remaining MHV amplitude from
\be
A(1^{-}, 3^{+}, 2^{-}, 4^{+}, 5^{+}) = \frac{s_{35}}{s_{13}}  A(1^{-}, 2^{-}, 4^{+}, 3^{+}, 5^{+}) + 
  \frac{s_{34} + s_{35}}{s_{13}} A(1^{-}, 2^{-}, 3^{+}, 4^{+}, 5^{+}) \,.
\ee

The five-graviton amplitude can now be obtained from the KLT formula
\begin{align}
\! M^{\rm CG}(1^{--} \! , 2^{--} \!, 3^{++} \!,  4^{++} \!,  5^{++}) &= i s_{13} s_{24} A^{(DF)^2}(1^-, 3^+, 2^-, 4^+, 5^+)  A^{\rm YM}(3^+,1^-,4^+, 2^-, 5^+)  \nn  \\
& \, \null + 
   i s_{12} s_{34}  A^{(DF)^2}(1^-, 2^-, 3^+, 4^+, 5^+)  A^{\rm YM}(2^-,1^-,4^+, 3^+,5^+) \,.  \nn \\
\end{align}
Comparing the the known result for conformal gravity MHV tree amplitudes, we see that it matches the compact result of Berkovits and Witten~\cite{Berkovits:2004jj},
\begin{align}
M^{\rm CG}(1^{--}  , 2^{--} , 3^{++},  4^{++} ,  5^{++}) \! \! \! &&=-i \spa{1}.{2}^4 \Big(\frac{\spa{1}.{2}^2 \spb{2}.{3}}{\spa{1}.{3}^2 \spa{2}.{3}}
+ \frac{\spa{1}.{4}^2 \spb{3}.{4}}{\spa{1}.{3}^2 \spa{3}.{4}}
   + \frac{\spa{1}.{5}^2 \spb{3}.{5}}{\spa{1}.{3}^2 \spa{3}.{5}}\Big) \nn \\ &&
   \times \Big(\frac{ \spa{1}.{2}^2 \spb{2}.{5}}{\spa{1}.{5}^2 \spa{2}.{5}}
   + \frac{\spa{1}.{3}^2 \spb{3}.{5}}{\spa{1}.{5}^2 \spa{3}.{5}} 
   + \frac{\spa{1}.{4}^2 \spb{4}.{5}}{\spa{1}.{5}^2 \spa{4}.{5}} \Big)\nn \\ &&
   \times \Big(\frac{\spa{1}.{2}^2 \spb{2}.{4}}{\spa{1}.{4}^2 \spa{2}.{4}} 
   + \frac{\spa{1}.{3}^2 \spb{3}.{4}}{\spa{1}.{4}^2 \spa{3}.{4}} 
   + \frac{\spa{1}.{5}^2 \spb{4}.{5}}{\spa{1}.{4}^2 \spa{4}.{5}}\Big)
\end{align}

The general formula for MHV superamplitudes in conformal supergravity is~\cite{Berkovits:2004jj}
\be
M^{\rm CSG}(1^-,2^-,3^+,\ldots,n^+) = (-1)^n \, i  \delta^{(8)}(Q) \prod_{i=3}^n \mathop{\sum_{j=1}}_{\! j\neq i}^{n} \frac{\spb{i}.{j}\spa{j}.{q}^2}{\spa{i}.{j}\spa{i}.{q}^2}
\ee
where the $\pm$ refers to the chirality of the graviton multiplet.  By explicit computation up to eight points, we have checked that the double copy between the dimension-six gauge theory and (super)-Yang-Mills agrees with this formula for MHV amplitudes.

\section{Deformations and Extensions}
\label{deforms}

There are several deformations or extensions that can be done to the dimension-six gauge theory (\ref{FullLagrDF}), while still maintaining color-kinematics duality. The fact that the duality holds means that these deformations directly correspond to extensions of the conformal gravity theory through the double copy. 

\subsection{Deforming with Mass Term: $(DF)^2 + {\rm YM}$ Gauge Theory }

Without changing the field content of the dimension-six theory, we can consider a one-parameter family of theories obtained by introducing a Yang-Mills term, $F^2$, to the Lagrangian (\ref{FullLagrDF}). Since this is a dimension-four operator we multiply it by a parameter of dimension two, $m^2$, which we can think of as a squared mass. This term will introduce a mass splitting between the physical gluon its corresponding ghost, the latter becoming massive, similar to the propagator in \eqn{mGhost}. However, the resulting theory does not obey color-kinematics duality unless we also give the scalar $\varphi^\alpha$ the same mass. 

The mass-deformed Lagrangian is then
\bea
{\cal L}_{(DF)^2 + {\rm YM}}&=& \frac{1}{2}(D_{\mu} F^{a\, \mu \nu})^2  - \frac{1}{3} g  F^3+ \frac{1}{2}(D_{\mu} \varphi^{\alpha})^2  + \frac{1}{2} g \,  C^{\alpha ab}  \varphi^{ \alpha}   F_{\mu \nu}^a F^{b\, \mu \nu }  +  \frac{1}{3!} g \, d^{\alpha \beta \gamma}   \varphi^{ \alpha}  \varphi^{ \beta} \varphi^{ \gamma} \nn \\ &&
\null  -   \frac{1}{2} m^2 (\varphi^{\alpha})^2- \frac{1}{4} m^2 (F^a_{\mu \nu})^2\,.
\label{massdefL}
\eea
By computing tree amplitudes up to eight points for external physical (massless) gluons, we have confirmed that this theory obeys color-kinematics duality (BCJ relations) for any value of the mass $m$. 

We can think of the Lagrangian as describing a $(DF)^2 + {\rm YM}$ theory that interpolates between marginal theories in $D=4$ and $D=6$ dimensions, and where $m$ plays the role of the interpolation parameter. Yang-Mills theory is recovered in the $m\rightarrow \infty$ limit, and $(DF)^2$ in the $m\rightarrow 0$ limit.

\subsection{Double Copy for Weyl-Einstein Gravity}
\label{sec:WEG}

The double copy between the mass-defomed theory (\ref{massdefL}) and Yang-Mills theory will give valid gravity amplitudes, even though we will not introduce any corresponding masses on the Yang-Mills side. This is guaranteed by the general consistency of color-kinematics duality. 

The resulting gravity theory will behave as follows: in the $m\rightarrow 0$ limit we will recover the non-minimal conformal gravity theory that we have already considered, and in the $m\rightarrow \infty$ limit the double copy will reduce to the standard double copy for non-pure Einstein gravity (including dilaton and axion scalars). Thus the gravity theory is a non-pure non-minimal version of Weyl-Einstein gravity. Again, the mass acts as the interpolation parameter between the two separate theories. 

In general, when we include supersymmetry on the YM side, we obtain the following four supergravity theories through the double copy:
\be
\big({\cal N}=0,1,2,4~\text{Weyl-Einstein SG}\big)=\big((DF)^2+{\rm YM}\big)\otimes \big({\cal N}=0,1,2,4~{\rm SYM}\big)\,,
\label{CG_DCsusy2}
\ee
where the mass $m$ is proportional to the Planck mass.

In the ${\cal N}=4$ case, it would be interesting to calculate the coefficient of the one-loop conformal anomaly (more specifically in the $m\rightarrow0$ limit, where the classical theory is conformal). The behavior of this potential anomaly may give support to the existence of theories that are ultraviolet finite to all orders, with a low-energy behavior that is indistinguishable from Einstein supergravities. Even so, the problem with ghosts would still remain.

\subsection{Adding Scalars to $(DF)^2$: Maxwell-Weyl Gravity }
\label{sec:MW}

The dimension-six gauge theory can be extended by adding to the spectrum $N$ scalars, $\phi^{aA}$, that transform in the adjoint of the gauge group, and in some real $N$-fold representation (e.g. adjoint) of a global group $G$. This has to done manually and not by dimensional reduction. As already mentioned in \sec{dim6Lsec}, dimensional reduction of the dimension-six theory does not yield interacting scalars. 

The kinetic term for the scalars is of the usual two-derivative type, and the extended dimension-six Lagrangian is then
\bea
{\cal L}_{(DF)^2+{\rm scalars}}&=&\frac{1}{2}(D_{\mu} F^{\mu \nu a})^2 - \frac{1}{3}  F^3 + \frac{1}{2}(D_{\mu} \varphi^{\alpha})^2 + \frac{1}{2}  g d^{ab \alpha}   F^{\mu \nu a} F_{\mu \nu }^b \varphi^{ \alpha}  + \frac{1}{3!} g d^{\alpha \beta \gamma}   \varphi^{ \alpha}  \varphi^{ \beta} \varphi^{ \gamma} \nn \\ &&
\null  + \frac{1}{2}(D_{\mu} \phi^{a A})^2  + \frac{1}{2} g d^{ab \alpha}   \phi^{a A} \phi^{b A} \varphi^{ \alpha}\,,
\label{dim6plusScalars} 
\eea
where the last term is a non-standard coupling to the scalar $\varphi^\alpha$. This term needs to be included if the theory is to obey color-kinematics duality. Computing amplitudes up to eight points for the external physical gluons or the adjoint scalars (i.e. external adjoint particles), again confirms that the theory obeys the duality. Other than the gauge coupling constant, the number of scalars and the gauge-group rank, there are no free parameters allowed by the duality (with the above limited choice of operators). 

The double copy between the extended theory (\ref{dim6plusScalars}) and (super-)Yang-Mills will include $N$ abelian vectors, thus the (super)gravity theory will be of Maxwell-Weyl type,
\be
\big({\cal N}=0,1,2,4~\text{Maxwell-Weyl SG}\big)=\big((DF)^2+{\rm scalars}\big)\otimes \big({\cal N}=0,1,2,4~{\rm SYM}\big)\,,
\label{CG_DCsusy3} 
\ee
where, as before, the theories are non-minimal conformal (super)gravities.

\subsection{Adding $\phi^3$ Interactions to $(DF)^2$:  Yang-Mills-Weyl Gravity}
\label{sec:YMW2}
Starting from the double copy (\ref{CG_DCsusy3}), we can immediately gauge the global symmetry group that the $N$ vectors transforms under, giving Yang-Mills-Weyl theories. The idea is the same as for the double-copy construction of Yang-Mills-Einstein supergravities used in ref.~\cite{Chiodaroli:2014xia}. We introduce a cubic self-interaction between the scalars in the extended dimension-six theory (\ref{dim6plusScalars}). The $\phi^3$ interaction is weighted by structure constants coming from both the gauge group, $f^{abc}$, and the global group, $\tilde{f}^{ABC}$, and it is also weighted by the gauge coupling and a free parameter $\lambda$,
\bea
{\cal L}_{(DF)^2+\phi^3}&=& \frac{1}{2}(D_{\mu} F^{a\, \mu \nu})^2  - \frac{1}{3} g  F^3+ \frac{1}{2}(D_{\mu} \varphi^{\alpha})^2  + \frac{1}{2} g \,  C^{\alpha ab}  \varphi^{ \alpha}   F_{\mu \nu}^a F^{b\, \mu \nu }  +  \frac{1}{3!} g \, d^{\alpha \beta \gamma}   \varphi^{ \alpha}  \varphi^{ \beta} \varphi^{ \gamma} \nn \\ &&
\null  + \frac{1}{2}(D_{\mu} \phi^{a A})^2 +  \frac{1}{2} g \, C^{\alpha ab}  \varphi^{ \alpha}  \phi^{a A} \phi^{b A} + \frac{1}{3!} g \lambda \, f^{abc} \tilde{f}^{ABC}  \phi^{a A} \phi^{b B} \phi^{c C}\,.
\label{dim6plusphi3}
\eea

This theory obeys color-kinematics duality for any value of the coupling $\lambda$, as has been checked by explicitly computing the tree amplitudes for external physical gluons and external adjoint scalars up to eight points. 

The double copy between the theory (\ref{dim6plusphi3}) and (super-)Yang-Mills now include a non-abelian Yang-Mills sector,
\be
\big({\cal N}=0,1,2,4~\text{Yang-Mills-Weyl SG}\big)=\big((DF)^2+\phi^3\big)\otimes \big({\cal N}=0,1,2,4~{\rm SYM}\big)\,,
\label{CG_DCsusy4}
\ee
The global group in (\ref{dim6plusphi3}) becomes promoted to the gauge group $G$, and similarly the free parameter $\lambda$ becomes promoted to the gauge coupling, in the Yang-Mills-Weyl supergravity theories. See ref.~\cite{Chiodaroli:2014xia} for further details in the analogous Yang-Mills-Einstein case.

\subsection{Combining Deformations: Yang-Mills-Weyl-Einstein Gravity}
\label{sec:YMEW2}

Finally, we can combine all the deformations and extensions of the dimension-six theory. The Lagrangian is
\bea
{\cal L}_{(DF)^2+{\rm YM}+\phi^3}&=& \frac{1}{2}(D_{\mu} F^{a\, \mu \nu})^2  - \frac{1}{3} g  F^3+ \frac{1}{2}(D_{\mu} \varphi^{\alpha})^2  + \frac{1}{2} g \,  C^{\alpha ab}  \varphi^{ \alpha}   F_{\mu \nu}^a F^{b\, \mu \nu }  \nn \\ &&
\null  +  \frac{1}{3!} g \, d^{\alpha \beta \gamma}   \varphi^{ \alpha}  \varphi^{ \beta} \varphi^{ \gamma}   + \frac{1}{2}(D_{\mu} \phi^{a A})^2  +  \frac{1}{2} g \, C^{\alpha ab}  \varphi^{ \alpha}  \phi^{a A} \phi^{b A} \nn \\
&& \null + \frac{1}{3!} g \lambda \, f^{abc} \tilde{f}^{ABC}  \phi^{a A} \phi^{b B} \phi^{c C}  -   \frac{1}{2} m^2 (\varphi^{\alpha})^2- \frac{1}{4} m^2 (F^a_{\mu \nu})^2\,.
\label{dim6plusYMplusphi3}
\eea
Remarkably, the theory still obeys color-kinematics duality for any value of the mass $m$ and  coupling $\lambda$, as we have checked up to eight points at tree level. 

The theory interpolates between the $(DF)^2+\phi^3$ theory (\ref{dim6plusphi3}) in the $m \rightarrow 0$ limit, and ${\rm YM}+\phi^3$ theory in the $m\rightarrow \infty$ limit. The latter theory was first described in ref.~\cite{Chiodaroli:2014xia}. Interestingly, the ${\rm YM}+\phi^3$ theory has a quartic interaction between the $\phi^{a A}$ scalars, which is not visible in the Lagrangian (\ref{dim6plusYMplusphi3}). It turns out that this quartic term only appears in the $m\rightarrow \infty$ limit after integrating out the infinitely massive $\varphi^{ \alpha}$, which now has become an auxiliary field. 

The double copy between the theory (\ref{dim6plusYMplusphi3}) and (super-)Yang-Mills gives amplitudes in Yang-Mills-Weyl-Einstein (super)gravities,
\be
\big({\cal N}=0,1,2,4~\text{Yang-Mills-Weyl-Einstein SG}\big)=\big((DF)^2+{\rm YM}+\phi^3\big)\otimes \big({\cal N}=0,1,2,4~{\rm SYM}\big)\,,
\label{CG_DCsusy5}
\ee
where, in the gravity theory, the parameters $m$ and $\lambda$ are proportional to the Planck mass and gauge coupling constant, respectively. Similar to the maximally supersymmetric Weyl-Einstein theory, it would be interesting to work out the coefficient of the one-loop conformal anomaly for the ${\cal N}=4$ Yang-Mills-Weyl-Einstein theory.

\section{Conclusions}
\label{sec:conclusion}

In this paper, we have studied amplitudes in conformal gravity theories obtained from double copies between (super-)Yang-Mills theories and a new bosonic gauge theory that is built entirely out dimension-six operators (using $D=6$ counting). The new gauge theory was uniquely determined using an ansatz consisting of five gauge-invariant operators and imposing that the resulting scattering amplitudes obey color-kinematics duality. By construction, the gauge theory is marginal, or classically conformal, in six dimensions. Related to this, it has a four-derivative kinetic term with associated $1/p^4$ propagators and ghost states. The appearance of ghosts in the gauge theory not surprising given that conformal gravity has precisely such states. 

By general consistency of the duality between color and kinematics~\cite{Bern:2008qj,Bern:2010ue}, one is guaranteed to obtain scattering amplitudes in some gravitational theory (i.e. a diffeomorphism invariant theory with massless spin-2 states) from the double copy between two duality-satisfying gauge theories~\cite{Chiodaroli:2017ngp}. By dimensional arguments the double copy between the dimension-six theory and ${\cal N}=0,1,2,4$ (super-)Yang-Mills give amplitudes in (super)gravity theories built out of curvature squared invariants, where the amount of supersymmetry is inherited from the Yang-Mills theory. Theories of this type include conformal supergravities, and explicit amplitude calculations confirm that this is indeed the class of theories given by the double copy.

Specifically, the obtained conformal gravity theories are of the non-minimal type, implying that there are non-trivial scalar functions multiplying the curvature squared invariants. (The minimal conformal (super)gravities have a trivial S-matrix for physical external states in four-dimensional flat space~\cite{Maldacena:2011mk,Adamo:2013tja,Adamo:2016ple}.) The non-minimal theories are perhaps most familiar in the context of ${\cal N}=4$ conformal supergravities where the distinction is important in the analysis of potential conformal and global anomalies~\cite{Fradkin:1983tg,Fradkin:1985am,Romer:1985yg}. Maximally supersymmetric conformal gravity is obtained by the double copy between ${\cal N}=4$ SYM and the new dimension-six gauge theory. The amplitudes that this construction gives agrees with the Berkovits-Witten conformal gravity twistor string~\cite{Berkovits:2004jj} (which is a closed subsector of Witten's original twistor string~\cite{Witten:2003nn}). We have checked this explicitly up to eight points in the MHV sector. It would be interesting to also check this in more complicated N$^k$MHV helicity sectors, and for amplitudes with non-physical ghost states on the external legs (e.g. see calculations in ref.~\cite{Dolan:2008gc}).

The bosonic dimension-six gauge theory was constructed out of dimension-agnostic operators, and thus it can be used to compute amplitudes in any spacetime dimension. The conformal supergravity double copy is then valid in any dimension where a corresponding super-Yang-Mills theory exists. Ten-dimensional  ${\cal N}=1$ conformal supergravity is an especially interesting case, since a unique action has been constructed for it a long time ago by de Roo~\cite{deRoo:1991at}. Given that the double copy computes ten-dimensional amplitudes in a curvature-squared theory, with supersymmetry manifestly inherited from ten-dimensional SYM, one can expect that double copy is uniquely identified with the theory in ref.~\cite{deRoo:1991at}. It would be interesting to confirm this by explicit calculations using the known action.

The new dimension-six gauge theory admits several deformations that respects the color-kinematics duality. We considered the addition of an arbitrary number of adjoint scalars. The needed couplings involve operators beyond the covariantized kinetic terms; however, all relative parameters are uniquely fixed by the duality. The resulting double copy with super-Yang-Mills theories give amplitudes in (non-minimal) Maxwell-Weyl conformal supergravities, where the number of photons are determined by the number of scalars in the underlying gauge theory. The global symmetry of the photons can be gauged by introducing $\lambda \phi^3$ self-interactions in the dimension-six gauge theory, where $\lambda$ is unconstrained by the duality. The resulting double copy with super-Yang-Mills theories give amplitudes in (non-minimal) Yang-Mills-Weyl conformal supergravities, where $\lambda$ plays the role of the non-abelian coupling constant. This construction is analogous to the double copy for Yang-Mills-Einstein supergravity~\cite{Chiodaroli:2014xia}.

The dimension-six gauge theory can also be deformed by a regular Yang-Mills term, $m^2 F^2$, which gives a mass to the gluon ghost state, separating it from the massless physical gluon. An additional mass term is needed for a scalar that appears in the gauge theory, in order for the duality to be preserved under this deformation. This results in a one-parameter family of duality satisfying gauge theories, interpolating between the dimension-six theory and the dimension-four Yang-Mills theory. The double copy of this theory with super-Yang-Mills theory gives a supergravity theory that interpolates between four-derivative conformal gravity and two-derivative Einstein gravity. 

Finally, combining all the deformations, the duality still holds and the double copy gives amplitudes in Yang-Mills-Weyl-Einstein supergravities. If it would not be for the ghost states in the Weyl sector, these theories would be very interesting from a phenomenological point of view, as they can be made to agree with standard model physics and Einstein gravity at low energies, and at the same time be have a behavior consistent with renormalizable or UV finite theories at high energies. Given that the ghost state of the Weyl sector can be traced back to the ghost states of the dimension-six gauge theory, one can have some hope that these states may be better understood in the much simpler framework of gauge theories. 

Beyond the four-derivative gravity theories considered so far, the double copy can also be applied to two copies of the dimension six gauge theory,
\be
(DF)^2 \otimes (DF)^2 \sim (\nabla {\cal R})^2 +{\cal R}^3\,,
\label{DC6}
\ee
which by dimensional counting must give rise to a bosonic six-derivative gravity theory that is marginal in six dimensions. The schematic form is indicated here (with ${\cal R}$ being the Riemann tensor or Ricci tensor/scalar), but other than recognizing the need for a kinetic term of the form $(\nabla {\cal R})^2$ and an interaction term of the form ${\cal R}^3$, no effort has gone into studying this theory. This theory has $1/p^6$ propagators, and thus the problem with ghost states is even more severe.

The duality between color and kinematics has been observed to also apply to the NLSM or chiral Lagrangian~\cite{Chen:2013fya,Cheung:2014dqa,Cachazo:2014xea,Du:2016tbc,Cheung:2016prv,Carrasco:2016ldy}. A number of different double copies has been constructed using the NLSM amplitudes and numerators. Given the increased arsenal of five new gauge theories that obeys color-kinematics duality: i.e the theories in eqs. (\ref{FullLagrDF}), (\ref{massdefL}), (\ref{dim6plusphi3}), (\ref{dim6plusScalars})  and (\ref{dim6plusYMplusphi3}), we can obtain various new double copies with the NLSM amplitudes.  The simplest one is obtained using the undeformed dimension-six theory (\ref{FullLagrDF}), giving a schematic double copy involving Einstein gravity and higher-derivative photon terms
\be
(DF)^2 \otimes {\rm NLSM} \sim \text{Einstein gravity}+(\nabla F)^2+\ldots
\label{DC6}
\ee
A more detailed Lagrangian for this theory is given in ref.~\cite{Azevedo:2017lkz}.

We conclude by noting that in this work significant progress has been made in understanding the double copy structure of conformal (super)gravity theories, expressing the amplitudes as a product of contributions coming from simpler-to-understand gauge theories. Given the deep connection between superconformal theories and ultraviolet finiteness, we hope that the current work will shed some light on the problem of potential ultraviolet divergences in supergravity.

\acknowledgments

We thank Tim Adamo, Johannes Broedel, Paolo Di Vecchia, Yvonne Geyer, Oliver Schlotterer, and Arkady Tseytlin for stimulating discussions on conformal gravity and related theories. We also thank Thales Azevedo, Marco Chiodaroli, Murat G\"{u}naydin, and Radu Roiban for collaborations on topics closely related to this work.  H.J. would like to thank the Kavli Institute for Theoretical Physics in Santa Barbara for hospitality during the program ``Scattering Amplitudes and Beyond", where part of this work was completed. This research is supported in part by the Swedish Research Council under grant 621-2014-5722, the Knut and Alice Wallenberg Foundation under grant KAW 2013.0235, the Ragnar S\"{o}derberg Foundation under grant S1/16, and by the National Science Foundation under grant NSF PHY-1125915.

\appendix

\section{Feynman Rules}
\label{FeynRules}

\newcommand{\vcenterincludegraphics}[2][]{%
  \begin{array}{@{}c@{}}
  \includegraphics[#1]{#2}
  \end{array}%
}
\def\m#1{\mu_{#1}}
\def\p#1{p_{#1}}
\def\a#1{a_{#1}}
\def\alph#1{\alpha_{#1}}
\def\cg{C}

Below we give the Feynman rules for the undeformed dimension-six theory (\ref{FullLagrDF}); for completeness we give the Lagrangian here again
\be
{\cal L}_{(DF)^2}=   \frac{1}{2}(D_{\mu} F^{a\, \mu \nu})^2- \frac{1}{3} g  F^3  +\frac{1}{2}(D_{\mu} \varphi^{\alpha})^2 + \frac{1}{2}g \,  C^{\alpha ab}  \varphi^{ \alpha}   F_{\mu \nu}^a F^{b\, \mu \nu }  + \frac{1}{3!} g \, d^{\alpha \beta \gamma}   \varphi^{ \alpha}  \varphi^{ \beta} \varphi^{ \gamma}\,.
\ee

We gauge-fix the Lagrangian by introducing a term $\frac{1}{2\xi}\left(\partial_{\mu}\partial_{\nu}A^{a\,\nu}\right)^{2}$, and then set $\xi=1$ to obtain the propagators in a Feynman-like gauge:

\hspace{2cm}

$\begin{aligned}
  \vcenterincludegraphics[width=.20\textwidth]{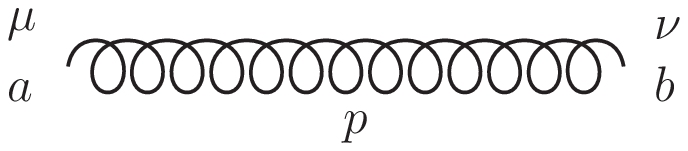} 
  \hspace{.5cm}=\hspace{.5cm}{}&
  \begin{aligned}
 i\frac{\eta^{\mu\nu}\delta^{ab}}{p^{4}}
  \end{aligned}
  {} \\[.5cm]
    \vcenterincludegraphics[width=.20\textwidth]{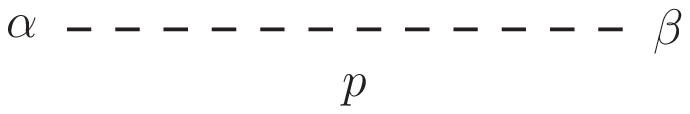} 
  \hspace{.5cm}=\hspace{.5cm}{}&
  \begin{aligned}
 i\frac{\delta^{\alpha\beta}}{p^{2}}
  \end{aligned}
  {} \\[.5cm]
\end{aligned}$

\noindent
As is conventional the gluons are represented by the curly lines, and the scalar $\varphi^\alpha$, in the auxiliary representation $R$, is marked by dashed lines. 

There are four distinct cubic vertices, three distinct quartic vertices, two quintic vertices and finally one sextic vertex:  

$\begin{aligned}
  \vcenterincludegraphics[width=.19\textwidth]{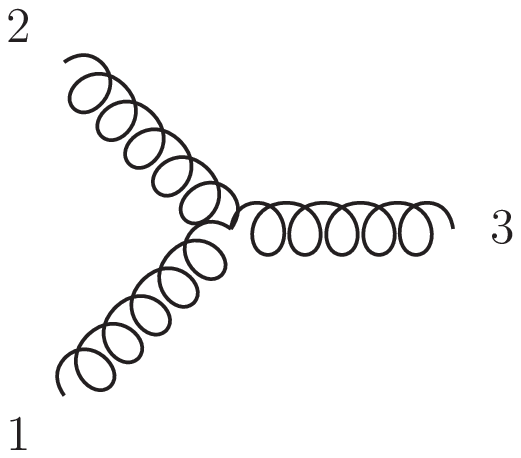} 
  \hspace{.5cm}=\hspace{.5cm}{}&
  \begin{aligned}
  g f^{\a1\a2\a3} \Bigl(&
  \p2\cdot\left(\p2+\p3\right) \p3^{\m3}\eta^{\m1\m2}
  - \left(\p1^{2} + (\p1+\p3)^{2}\right)\p2^{\m3}\eta^{\m1\m2}
  \\&
  + \p3^{\m1}\left(\tfrac{2}{3}\p1^{\m2}\p2^{\m3} + 2\p2^{\m2}\p2^{\m3} + \p2^{\m2}\p3^{\m3}\right)
  \Bigr) + \text{perm.}(1,2,3)
  \end{aligned}
  {} \\[.5cm]
  \vcenterincludegraphics[width=.19\textwidth]{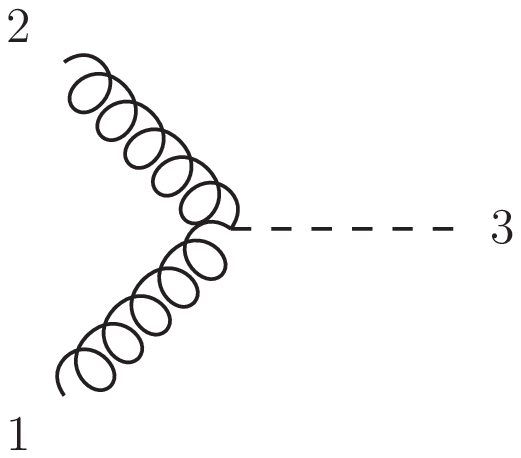} 
  \hspace{.5cm}=\hspace{.5cm}{}&
  \begin{aligned}
  -2 i g \cg^{\alph3 \a1\a2} (\p1\cdot\p2 \eta^{\m1\m2}- \p2^{\m1}\p1^{\m2})
  \end{aligned}
  {} \\[.5cm]
  \vcenterincludegraphics[width=.19\textwidth]{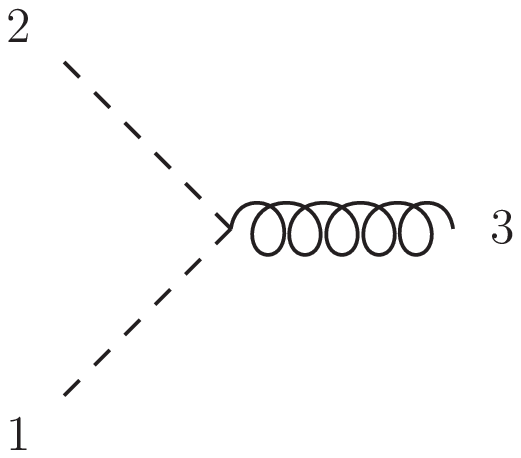} 
  \hspace{.5cm}=\hspace{.5cm}{}&
  \begin{aligned}
  ig(T_{R}^{\a3})^{\alph1\alph2}(\p1-\p2)^{\m3}
  \end{aligned}
  {} \\[.5cm]
  \vcenterincludegraphics[width=.19\textwidth]{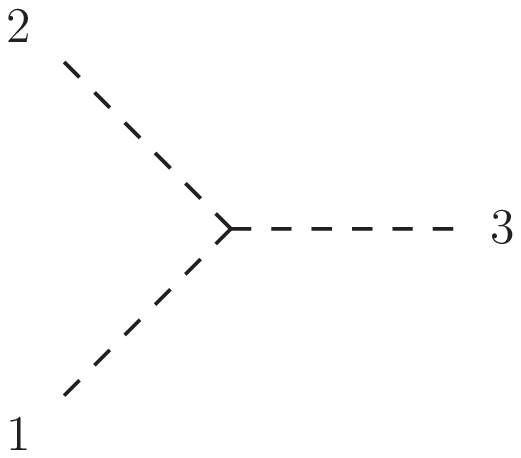} 
  \hspace{.5cm}=\hspace{.5cm}{}&
  \begin{aligned}
  i g d^{\alph1\alph2\alph3}
  \end{aligned}
  {} \\[.5cm]
\end{aligned}$

$\begin{aligned}
  \vcenterincludegraphics[width=.19\textwidth]{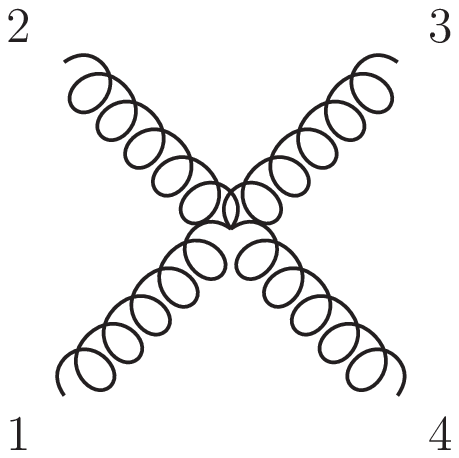} 
  \hspace{.5cm}=\hspace{.5cm}{}&
  \begin{aligned}
  & -i g^{2} f^{\a1\a2b}f^{b \a3\a4}
  \\ &\hspace{.5cm}
  \times\Bigl(
  \tfrac{1}{2} \p2\cdot \p4\eta^{\m1\m2}\eta^{\m3\m4} + (\p3 + \p4)\cdot\p4 \eta^{\m1\m4}\eta^{\m2\m3}
  \\ &\hspace{1cm}
  - \left(\tfrac{1}{2}\p1^{\m1}\p4^{\m4} + 2\p4^{\m1}\p3^{\m4} + 2\p2^{\m1}\p4^{\m4} + \p4^{\m1}\p4^{\m4}\right)\eta^{\m2\m3}
  \\ &\hspace{1cm}
  + \p4^{\m1} (2\p1 + \p2 + \p3)^{\m2} \eta^{\m3\m4} + 2\p2^{\m1}\p4^{\m3}\eta^{\m2\m4}
  \Bigr)
  \\ &\hspace{.5cm} 
  + \text{perm.}(1,2,3,4)
  \end{aligned}
  {} \\[.5cm]
   \vcenterincludegraphics[width=.19\textwidth]{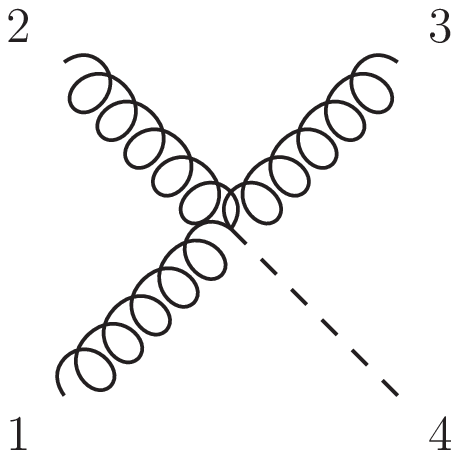} 
  \hspace{.5cm}=\hspace{.5cm}{}&
  \begin{aligned}
  -2 g^{2} f^{\a1\a2 b} \cg^{\alph4 b\a3} \p3^{\m1}\eta^{\m2\m3} + \text{perm.}(1,2,3)
  \end{aligned}
  {} \\[.5cm]
   \vcenterincludegraphics[width=.19\textwidth]{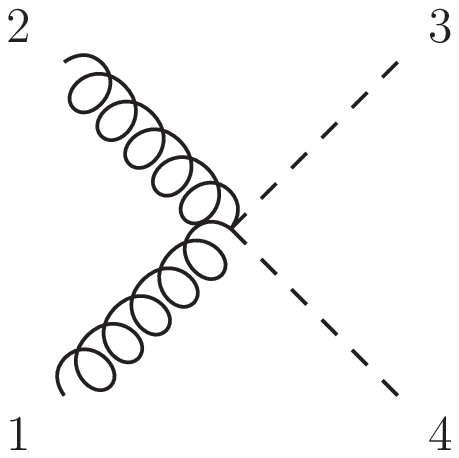} 
  \hspace{.5cm}=\hspace{.5cm}{}&
  \begin{aligned}
  -i g^{2} \left( (T_{R}^{\a1})^{\alph3\beta}(T_{R}^{\a2})^{\alph4\beta}
  + (T_{R}^{\a1})^{\alph4\beta}(T_{R}^{\a2})^{\alph3\beta}\right) \eta^{\m1\m2}
  \end{aligned}
  {} \\[.5cm]
\end{aligned}$  
  
$\begin{aligned}
  \vcenterincludegraphics[width=.19\textwidth]{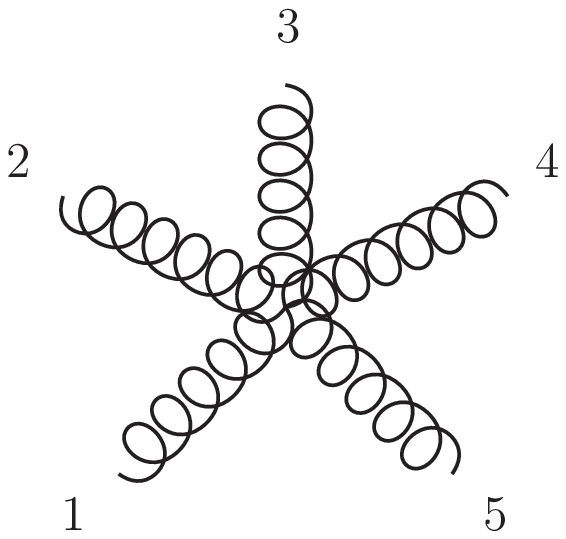} 
  \hspace{.5cm}=\hspace{.5cm}{}&
  \begin{aligned}
  & g^{3} f^{\a1\a2b}f^{b \a3 c} f^{c\a4\a5}
  \\ &\hspace{.5cm}
  \times\Bigl(
  \p2^{\m5}\eta^{\m1\m2}\eta^{\m3\m4}+(2\p3+2\p4+\p5)^{\m5}\eta^{\m1\m4}\eta^{\m2\m3}
  \Bigr)
  \\ &\hspace{.5cm} 
  + \text{perm.}(1,2,3,4,5)
  \end{aligned}
  {} \\[.5cm]
   \vcenterincludegraphics[width=.19\textwidth]{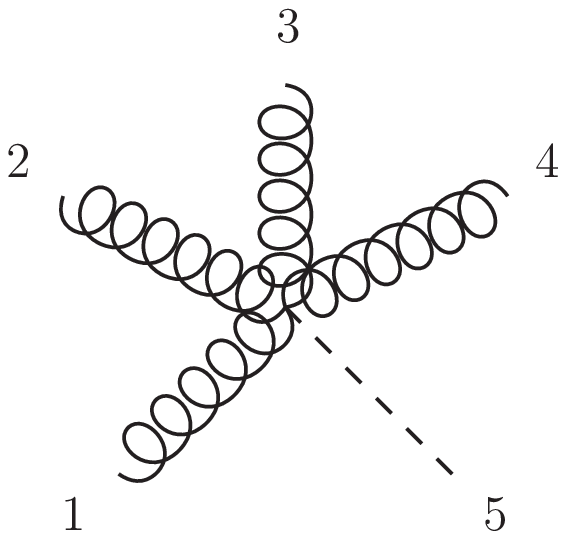} 
  \hspace{.5cm}=\hspace{.5cm}{}&
  \begin{aligned}
  & -\frac{1}{2} i g^{3} f^{\a1\a2 b}f^{\a3\a4 c} \cg^{\alph5 b c} \eta^{\m1\m4}\eta^{\m2\m3}
  + \text{perm.}(1,2,3,4)
  \end{aligned}
  {} \\[.5cm]
  %
  %
%
  %
  %
  \vcenterincludegraphics[width=.19\textwidth]{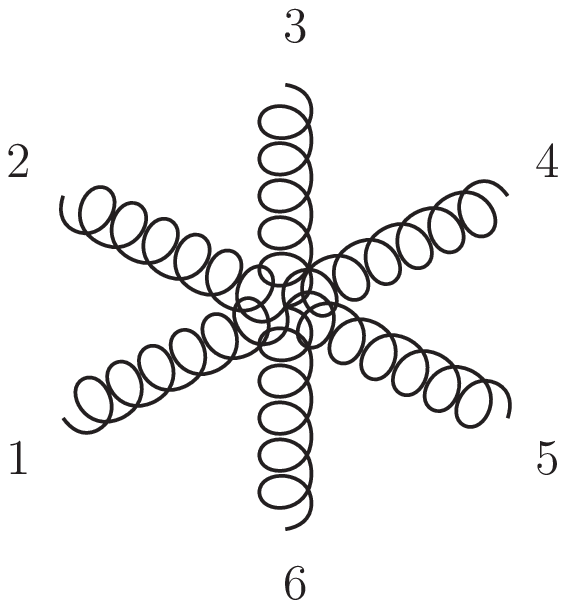} 
  \hspace{.5cm}=\hspace{.5cm}{}&
  \begin{aligned}
  i g^{4} &\Bigl(
  f^{\a1\a2b}f^{b \a3 c} f^{c\a4 d}f^{d\a5\a6}\eta^{\m1\m6}\eta^{\m2\m3}\eta^{\m4\m5}
  \\ &\hspace{.5cm}
  +  f^{\a1\a2b}f^{\a3\a4c} f^{\a5\a6 d}f^{bcd}\eta^{\m1\m6}\eta^{\m2\m4}\eta^{\m3\m5}
  \Bigr)
  \\ &
  + \text{perm.}(1,2,3,4,5,6)
  \end{aligned}
  {} \\[.5cm]
\end{aligned}$

\section{Pictorial Lie Algebra Relations}
\label{LieAlgebraSect}

In this appendix we give pictorial forms of the Lie-algebra relations that are used for reducing the tree-level color factors into a basis.

 In particular, if all $m$ external legs are transforming in the adjoint then these identities reduce the color factors to the usual $(m-2)!$ Kleiss-Kuijf/Del Duca-Dixon-Maltoni basis~\cite{Kleiss:1988ne,DelDuca:1999rs}, consisting of strings of contracted $f^{abc}$'s. If all but one of the external legs are in the adjoint, and the last leg is in the auxiliary representation $R$, the color factors can be reduced to at most one $(C^{\alpha})^{ab}$ tensor times a string of contracted $f^{abc}$'s. 
 
In order to compute loop level amplitudes in the dimension-six theory (\ref{FullLagrDF}) one also need to know how to evaluate the symmetric traces ${\rm STr}(T^{a_1}_R T^{a_2}_R \cdots T^{a_{2k}}_R)$; we leave this exercise to future work.

\begin{figure}[!htb]
\centering
\includegraphics[scale=.66]{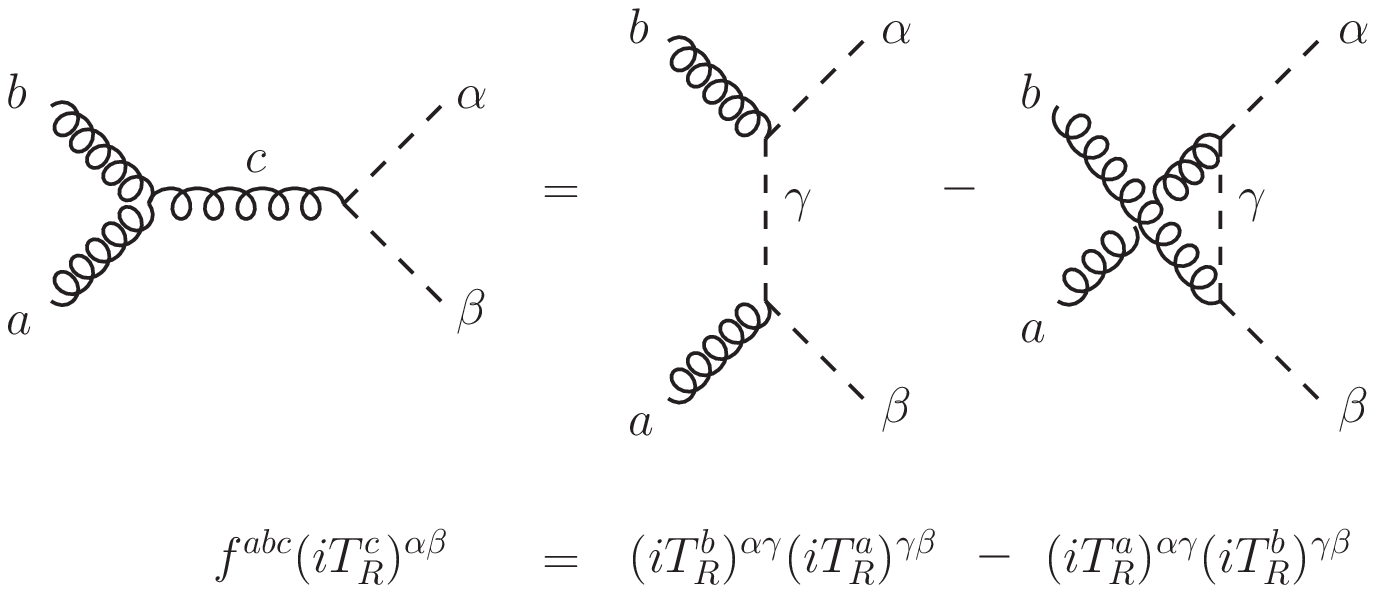}\\[0.5cm]
\includegraphics[scale=.66]{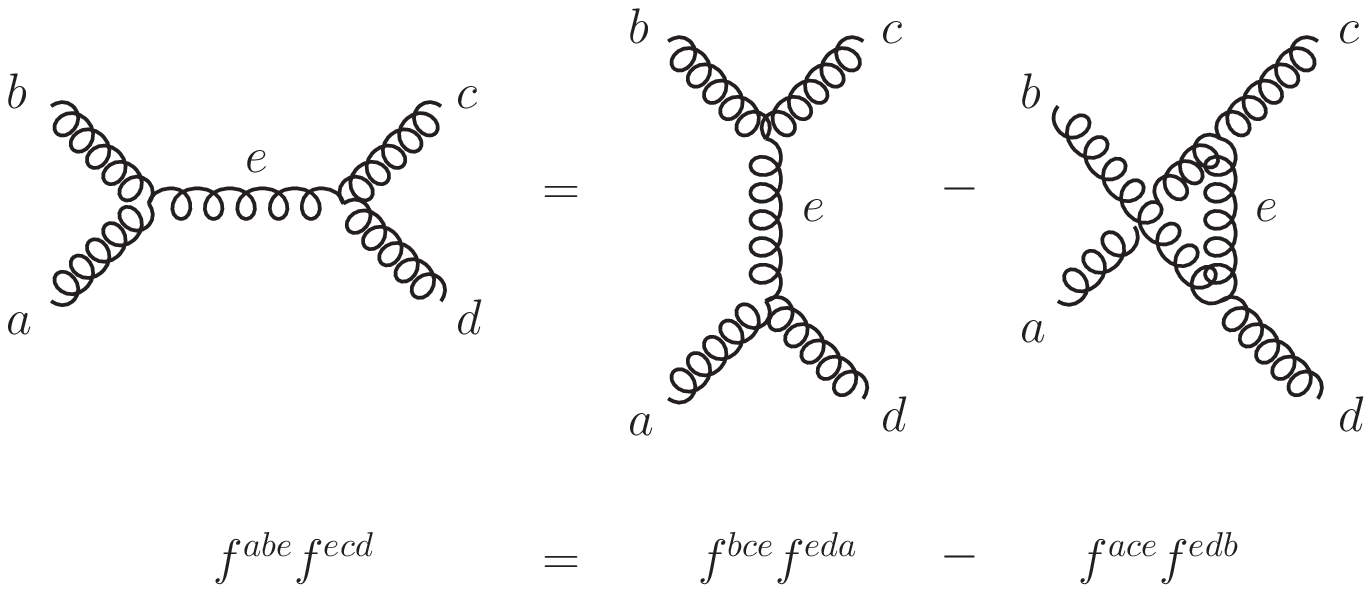}\\[0.5cm]
\includegraphics[scale=.66]{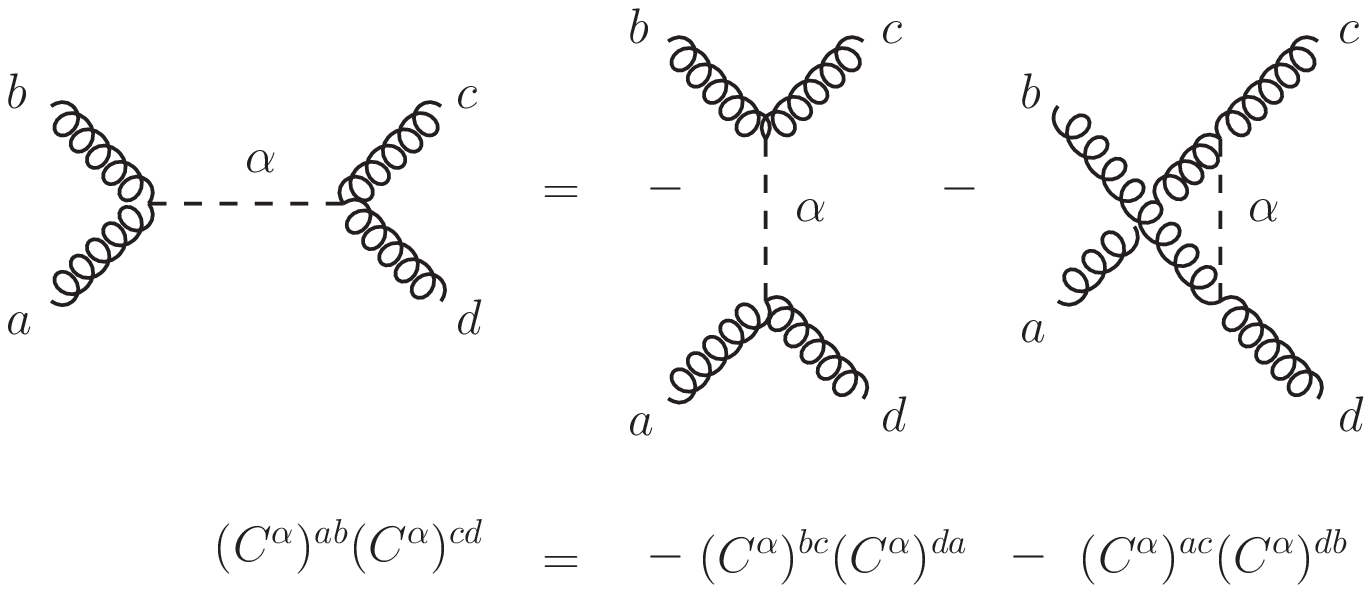}
\caption[a]{(a): The relation (\ref{1stID}) that enforces that $(T_R^a)^{\alpha \beta}$ is a covariant tensor under infinitesimal gauge-group rotations. (b): The Jacobi identity (enforces that $f^{abc}$ is a covariant tensor under infinitesimal gauge-group rotations). (c): A nice but redundant relation; it follows directly from the identity in \fig{GeneralizedJacobis}(a).
  }
\label{GeneralizedJacobis3}
\end{figure}

\begin{figure}[h]
\centering
\includegraphics[scale=.66]{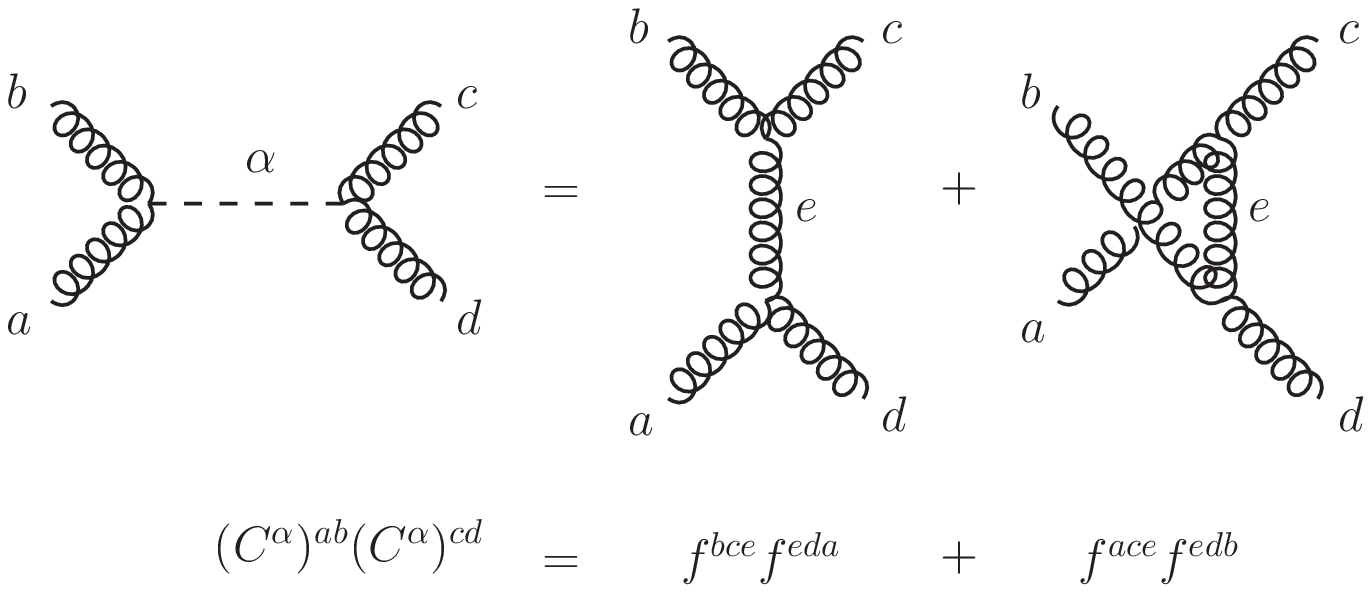}\\[0.5cm]
\includegraphics[scale=.66]{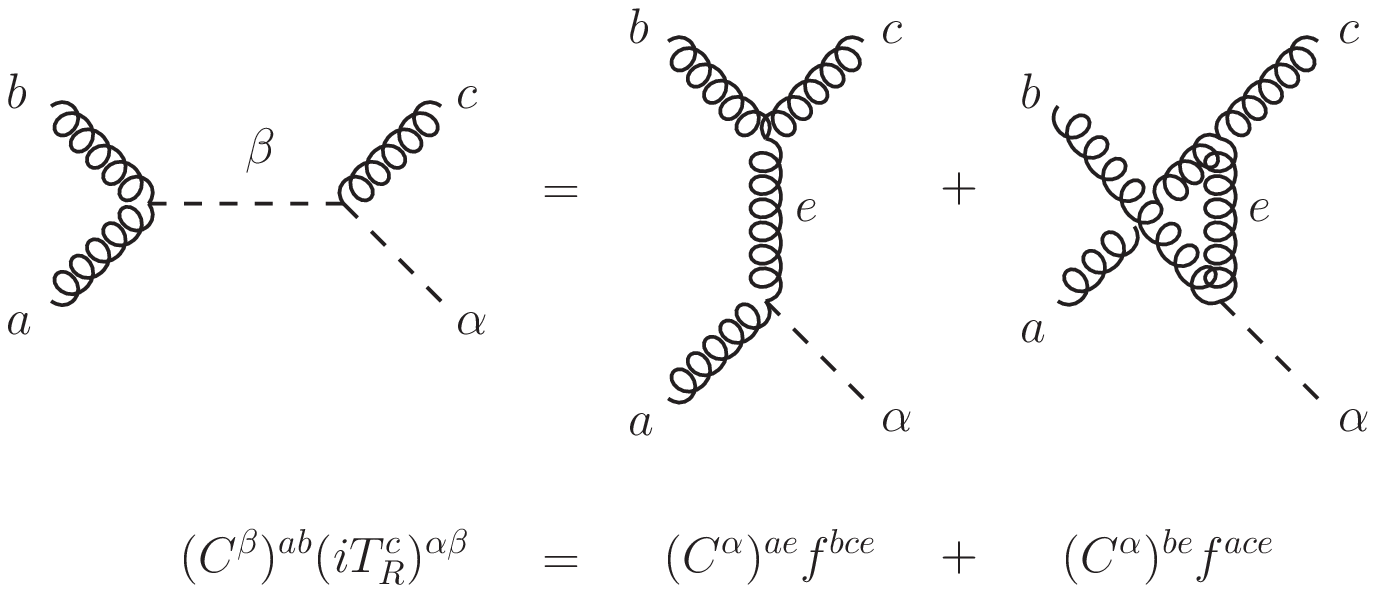}
\caption[a]{(a): The reduction relation (\ref{Id1a}) for contracting two $(C^{\alpha})^{ab}$ tensors. (b): The relation (\ref{2ndID}) that enforces that $(C^{\alpha})^{ab}$ is a covariant tensor under infinitesimal gauge-group rotations.}
\label{GeneralizedJacobis}
\end{figure}

\begin{figure}[!htb]
\centering
 \includegraphics[scale=.66]{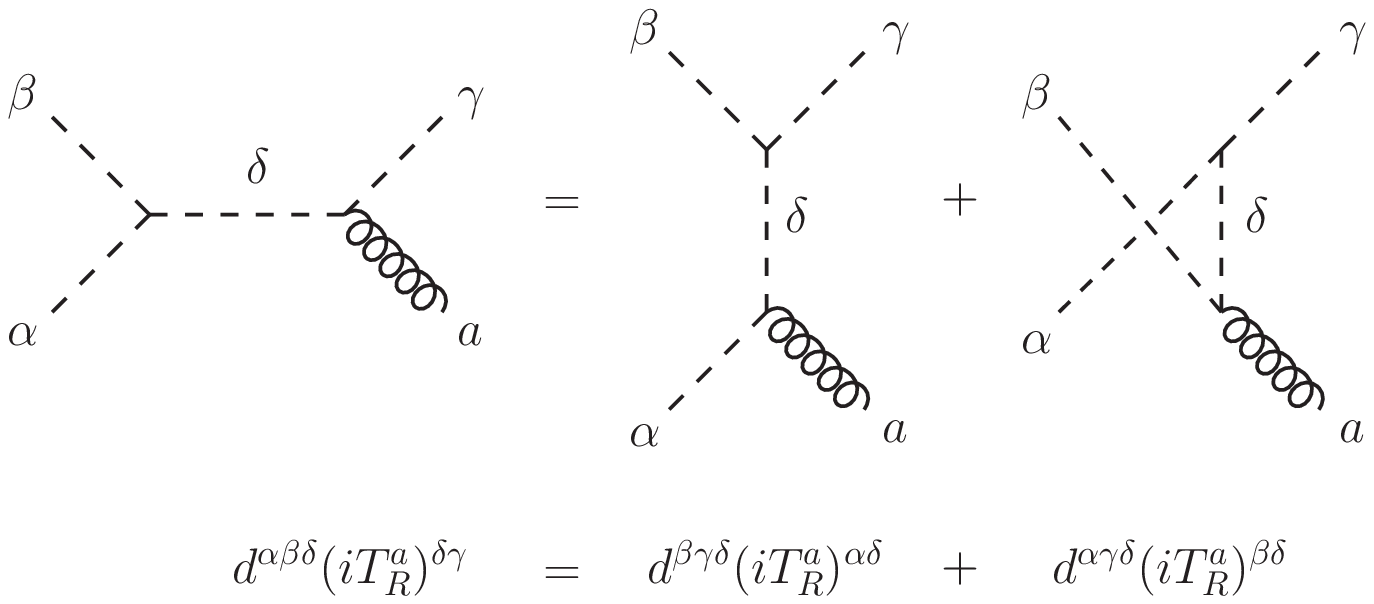}\\[0.5cm]
 \includegraphics[scale=.83]{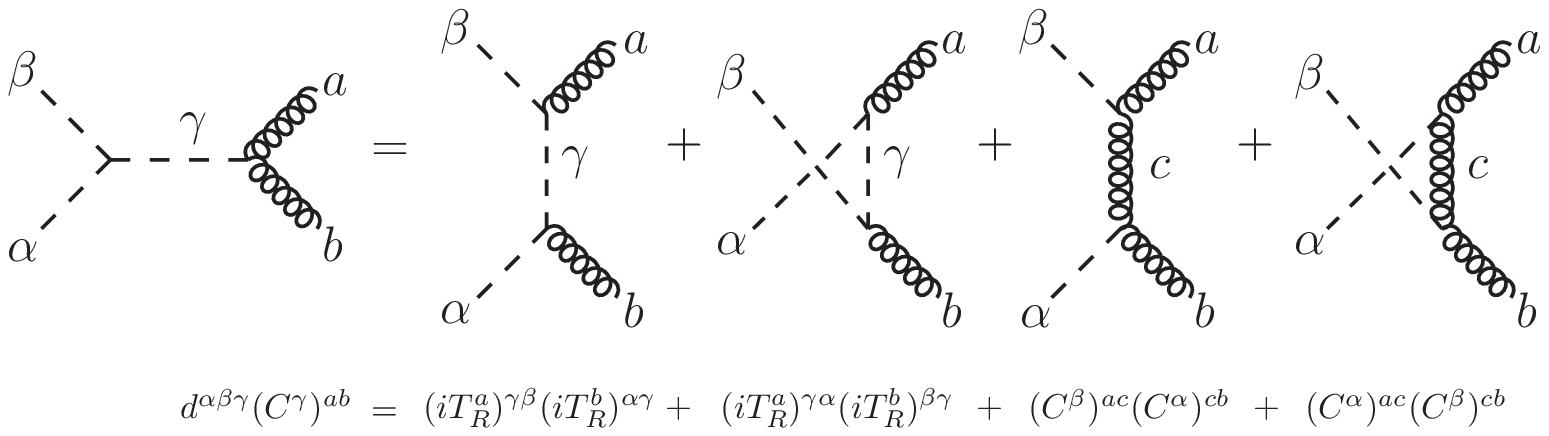}
\caption[a]{(a):  The relation (\ref{3rdID}) that enforces that $d^{\alpha \beta \gamma}$ is a covariant tensor under infinitesimal gauge-group rotations. (b): The reduction relation (\ref{Id1b})  for $d^{\alpha\beta\gamma}$ contracted with $(C^{\gamma})^{ab}$; it can be more compactly written in terms of anti-commutators: $d^{\alpha\beta\gamma}(C^{\gamma})^{ab} = \{i\Trep^{a},i\Trep^{b}\}^{\alpha\beta} +  \{C^{\alpha},C^{\beta}\}^{ab}$.
  }
\label{GeneralizedJacobis2}
\end{figure}

\bibliographystyle{JHEP}
\bibliography{references}

\end{document}